\documentclass[12pt,twoside]{article}
\usepackage{cmp2e}

\usepackage{graphicx}
\usepackage{amssymb}

\newcommand{\rd}{\mathrm{d}}
\newcommand{\ri}{\mathrm{i}}
\newcommand{\re}{\mathrm{e}}

\newcommand{\Imm}{\,\mathrm{Im}\,}

\title[\mathsurround=0pt%
Green's functions of infinite-$U$ Hubbard model]%
{\vspace{-2ex}%
\mathversion{bold}\mathsurround=0pt%
Green's functions of infinite-$U$ asymmetric Hubbard
model: Falicov-Kimball limit}

\author{I.V.Stasyuk, O.B.Hera}
\address{Institute for Condensed Matter Physics\\
of the National Academy of Sciences of Ukraine,\\
1 Svientsitskii Str., 79011 Lviv, Ukraine}

\date{Received February 24, 2003}
\begin{document}
\setcounter{page}{127}
\maketitle

\begin{abstract}
\mathsurround=0pt%
The asymmetric Hubbard model is used in investigating the lattice
gas of the moving particles of two types. The model is considered
within the dynamical mean-field method. The effective single-site
problem is formulated in terms of the auxiliary Fermi-field. To
solve the problem an approximate analytical method based on the
irreducible Green's function technique is used. This approach is
tested on the Falicov-Kimball limit (when the mobility of ions of
either type is infinitesimally small) of the infinite-$U$ case of
the model considered. The dependence of chemical potentials on
concentration is calculated using the one-particle Green's
functions, and different approximations are compared with the
exact results obtained thermodynamically. The densities of states
of localized particles are obtained for different temperatures and
particle concentrations. The phase transitions are investigated
for the case of the Falicov-Kimball limit in different
thermodynamic regimes.

\keywords asymmetric Hubbard model, Falicov-Kimball model, dynamical
mean-field theory, Green's functions, phase transitions

\pacs 71.10.Fd, 05.30.Fk, 05.70.Fh
\end{abstract}

\section{Introduction}

Strongly correlated electron systems have been a subject of
growing interest in recent years. These systems became especially
interesting after the discovery of high-$T_{\mathrm{c}}$
superconductivity. In general, the methods used for their
description are based on the Hubbard model and its generalizations
taking into account strong short-range correlations of particles.
Similar models are considered in investigating the ionic
conductivity in crystalline materials. Fermi lattice gas models
can be mentioned among them. The asymmetric Hubbard model
\cite{gruber} arises by extending the models to the case of the
systems with moving ions of two types. The transfer parameters and
chemical potentials are different for the ions of different
nature. As some special cases, the asymmetric Hubbard model
includes the Falicov-Kimball model and the standard Hubbard model.
Thereby, it should be mentioned that the asymmetric Hubbard model
has been recently obtained as a generalization of the electron
Falicov-Kimball model which describes the interactions between
itinerant $d$-electrons and localized $f$-electrons by including
$f$--$f$ hopping \cite{Batista}. The model was used in describing
the effects related to orbital ordering, as well as to
Bose-condensation of electron-hole pairs (which includes a
spontaneous ferroelectric polarization) in oxide compounds.

Despite the relative simplicity of these models, the theory of
electron spectrum and thermodynamic properties of such systems is
far from its final completion. In recent years the essential
achievements of the theory of strongly correlated electron systems
have been connected with the development of the dynamical mean-field
theory (DMFT) \cite{dens1}. This method is exact in the limit of
infinite space dimension ($d=\infty$) and is based on the fact
that the irreducible (according to Larkin) part of the
one-particle Green's function is single-sited (at appropriate
scaling of transfer parameter).

The central point in this method is the formulation and the
solution of the auxiliary single-site problem. In this case a
separated lattice site is considered as placed in some effective
environment. Since the processes of the particle hopping from the
site and returning into the site are taken into account, the mean
field acting on the particle states of the site is of a dynamic
nature. The field is described by the coherent potential
$J_{\sigma}(\omega)$ that should be determined in a
self-consistent way from the condition that the same total
irreducible (according to Larkin) part defines Green's functions
both in the atomic limit and for the lattice
\cite{dens2,Met1,Met2}:
\begin{eqnarray}
&&G_\sigma^{(a)}(\omega_n)=\frac{1}{\Xi_\sigma^{-1}(\omega_n)-J_\sigma(\omega_n)},
 \label{sys2} \\
&&G_\sigma (\omega_n,
\mathbf{k})=\frac{1}{\Xi_\sigma^{-1}(\omega_n)-t^\sigma_\mathbf{k}},
  \label{sys1} \\
&&G_\sigma^{(a)}(\omega_n)=G_{ii}^\sigma(\omega_n)=\frac{1}{N}\sum_\mathbf{k}
G^\sigma
  (\omega_n, \mathbf{k}).
  \label{sys3}
\end{eqnarray}
Here $\sigma$ is the type index (or a spin index for electron
systems); there is a possibility of the transfer parameter
$t^\sigma_\mathbf{k}$ can depend on a particle type. Summation over
the wave vector can be changed by integration with the density of
states. There are cases that are usually investigated in the
$d=\infty$ limit: the hypercubic lattice which is the
generalization of the cubic $d=3$ lattice, and the Bethe lattice
which is the thermodynamic limit of the Cayley tree when the
number of the nearest neighbours tends to infinity. The density of
states is Gaussian for a hypercubic lattice \cite{dens1}, while on
the Bethe lattice the density of states is semielliptic
\cite{dens2}.

The single-site problem can be solved analytically only in some
simple cases. In general, the application of numerical methods
turns out to be necessary. In the case of the Falicov-Kimball
model it is possible to investigate the thermodynamic properties
of the system by analytical calculation of the grand canonical
potential \cite{Freericks1,Falicov1}. Also, the analytical
expressions for the itinerant and  for the localized electron
Green's functions have been obtained \cite{brur,Freericks4}.
However, the topology of the localized electron band has not been
completely investigated yet.

Recently, an approximate analytical approach to the solving of a
single-site problem has been developed \cite{ista1,ista2}. This
approach is based on the irreducible Green's function technique
with projecting on the Hubbard basis of Fermi-operators. A system
of DMFT equations was obtained in the approximation which is a
generalization of the Hubbard-III approximation combined with a
self-consistent renormalization of the local electron levels (here
this approach will be called GH3). It has been shown that the
proposed approach includes, as special cases, a number of known
approximations (Hubbard-III, alloy-analogy~(AA) and modified
alloy-analogy~(MAA)).

In the present work this approach is formulated for the asymmetric
Hubbard model. Its applicability within different approximations
is tested on the infinite-$U$ spinless Falicov-Kimball model. The
densities of states of the moving and the localized particles are
obtained for  the Bethe lattice at different particle
concentrations and temperatures. Dependence of the chemical
potentials on concentration is calculated using the Green's
functions. The results of different approximations are compared
with the exact results obtained thermodynamically. Some
thermodynamic properties (such as phase transitions and phase
separations) of the system in different thermodynamic regimes are
investigated as well.

\section{Model}

We consider the case when the lattice gas model is used in
describing the systems with moving particles (ions or electrons)
of two types. A single site can be in three possible states:
(i)~no particles, (ii)~one particle of the first ($A$) type,
(iii)~one particle of the second ($B$) type. Motion of the
particles can be described by the creation and by the annihilation
operators and by the transfer parameters dependent on the particle
type. Since the site can be occupied by only one particle of
either type, the operators are not of the Fermi type. It is
possible to investigate the system using the Fermi operators by
formal adding the fourth state of a site occupied by two (A and B)
particles simultaneously. In this case, the asymmetric Hubbard
model arises. The Hamiltonian is
\begin{eqnarray}
&&  H=\sum_i H_i+\sum_{\langle ij \rangle} t_{ij}^A a^{A+}_i a_j^A
  +\sum_{\langle ij \rangle} t_{ij}^B a^{B+}_i a_j^B,
  \label{h1} \\
&& H_i=-\mu_A n_i^A-\mu_B n_i^B+U n^A_i n^B_i,
\end{eqnarray}
with chemical potentials dependent on the particle type. It can be
easily seen that Hamiltonian (\ref{h1}) corresponds to the
Hubbard model in an external magnetic field with the
spin-dependent electron transfer. The $U$ term is the single-site
interaction energy between the particles and it should tend to
infinity when we wish to return to the initial lattice gas model.

In the limit of large dimensions, the problem is investigated
using the DMFT approach. The single-site problem is formulated in
terms of the auxiliary Fermi-field \cite{ista1}. Let us write the
effective single-site Hamiltonian in the Hubbard operators
representation \arraycolsep=2pt
\begin{eqnarray}
  H_{\mathrm{eff}} & = & - \sum_{\sigma} \Big[  \mu_{\sigma} \big( X^{\sigma\sigma}+X^{22}
  \big) \Big] + U X^{22} \nonumber\\
   & & {}+ V \sum_{\sigma} \Big[ \big( X^{\sigma 0} + \zeta X^{2\bar{\sigma}} \big) \xi_{\sigma}
   + \xi^{+}_{\sigma} \big( X^{0\sigma} + \zeta X^{\bar{\sigma} 2}  \big)  \Big]
   +H_\xi\,,
\end{eqnarray}
where the notations are used: $\bar{\sigma}=B,\; \zeta=+$ for
$\sigma=A$; $\bar{\sigma}=A,\; \zeta=-$ for $\sigma=B$. The basis
of single-site states $| n_A, n_B\rangle$ is
\begin{equation}
\begin{array}{lll}
|0\rangle=|0,0\rangle , &  &\qquad |A\rangle=|1,0\rangle, \\
|2\rangle=|1,1\rangle, &  &\qquad |B\rangle=|0,1\rangle .
\end{array}
\label{BaseStates2}
\end{equation}
The Green's function $G_\sigma^{(a)}(\omega) \equiv \langle\langle
a_\sigma | a_\sigma^+ \rangle\rangle_\omega$ can be written in the
following form:
\begin{equation}
 G^{(a)}_{\sigma}=\langle\langle X^{0\sigma}| X^{\sigma 0} \rangle\rangle_{\omega}
 +\zeta \langle\langle X^{0\sigma}| X^{2\bar{\sigma}} \rangle\rangle_{\omega}
 +\zeta \langle\langle X^{\bar{\sigma}2}| X^{\sigma0} \rangle\rangle_{\omega}
 +\langle\langle X^{\bar{\sigma}2}| X^{2\bar{\sigma}} \rangle\rangle_{\omega}\,.
 \label{Grin_X}
\end{equation}
The auxiliary Fermi-field describes the environment of the
selected site and is formally characterized by the Hamiltonian
$H_\xi$. The explicit form of this Hamiltonian is unknown.
However, to calculate the Green's function
$G_\sigma^{(a)}(\omega)$, the averaging over the $\xi$, $\xi^+$
operators is done with the help of the function
\begin{equation}
  \mathcal{G}_\sigma
  (\omega)=\langle\langle \xi_\sigma|\xi_\sigma^+
  \rangle\rangle_\omega^{(H_\xi)}.
  \label{GrXi}
\end{equation}
The relation
\begin{equation}
  2\pi V^2   \mathcal{G}_\sigma(\omega)=J_\sigma (\omega)
\end{equation}
takes place in this case. Unlike the standard Hubbard model, the
coherent potential $J_\sigma$ is dependent on the type of the
particles ($J_A \neq J_B$ for $t^A_{ij} \neq t^B_{ij}$).

\section{Green's functions for the effective single-site problem}
Equations for functions (\ref{Grin_X}) are written using the
equations of motion for $X$-operators. According to the method
developed in \cite{Prog1,Prog2}, let us separate (in the Green's
functions of a higher order) the irreducible parts expressing the
derivatives $\ri\frac{\rd X^{0\sigma(\bar\sigma2)}}s{\rd
t}=[X^{0\sigma(\bar\sigma2)},H_{\mathrm{eff}}]$ as the sums of
regular (projected on the subspace formed by operators
$X^{0\sigma}$ and $X^{\bar\sigma 2}$) and irregular parts
\arraycolsep=2pt
\begin{eqnarray}
   \big[ X^{0\sigma}, H_{\mathrm{eff}}\big]& = &
   -\mu_\sigma X^{0\sigma}+ \alpha_1^{0\sigma}X^{0\sigma}+
   \alpha_2^{0\sigma} X^{\bar\sigma 2} + Z^{0\sigma},
   \nonumber \\
   \big[ X^{\bar{\sigma}2}, H_{\mathrm{eff}}\big] & = &
   (U-\mu_\sigma)X^{\bar\sigma 2}+\alpha_1^{\bar\sigma 2}
    X^{0\sigma}+\alpha_2^{\bar\sigma 2}X^{\bar\sigma 2} +Z^{\bar\sigma 2}.
\end{eqnarray}

Operators $Z^{0\sigma}$ and $Z^{\bar\sigma 2}$ are defined as
orthogonal to the operators from the basic subspace and we
come to the expressions
\begin{eqnarray}
  Z^{0\sigma} = V \overline{(X^{00}+X^{\sigma\sigma})\xi_\sigma}
  + V \overline {X^{\bar\sigma \sigma} \xi_{\bar\sigma}}
  +\zeta V \overline{X^{02}\xi^{+}_{\bar\sigma}},  \nonumber \\
  Z^{\bar\sigma 2}=V \overline{(X^{22}+X^{\bar\sigma \bar\sigma})\xi_\sigma}
  - \zeta V \overline {X^{\bar\sigma \sigma} \xi_{\bar\sigma}}
  - V \overline{X^{02}\xi^{+}_{\bar\sigma}},
\end{eqnarray}
where
\begin{eqnarray}
 \overline{(X^{00}+X^{\sigma\sigma})\xi_\sigma}&=&
 (X^{00}+X^{\sigma\sigma})\xi_\sigma,
 \nonumber \\
 \overline{(X^{22}+X^{\bar\sigma \bar\sigma})\xi_\sigma} &=&
 (X^{22}+X^{\bar\sigma \bar\sigma})\xi_\sigma,
 \nonumber \\
 \overline {X^{\bar\sigma \sigma} \xi_{\bar\sigma}} &=&
  X^{\bar\sigma \sigma} \xi_{\bar\sigma} -
  \frac{1}{A_{0\sigma}} \langle \xi_{\bar\sigma} X^{\bar\sigma 0}\rangle X^{0\sigma} -
  \frac{1}{A_{2 \bar\sigma}}  \langle X^{2\sigma} \xi_{\bar\sigma} \rangle X^{\bar\sigma 2} ,
 \nonumber \\
 \overline{X^{02} \xi^{+}_{\bar\sigma}}&=&
 X^{02} \xi^{+}_{\bar\sigma}-
 \frac{1}{A_{0\sigma}} \langle X^{\sigma 2} \xi_{\bar\sigma}^{+} \rangle X^{0\sigma} -
 \frac{1}{A_{2 \bar\sigma}} \langle \xi_{\bar\sigma}^{+} X^{0 \bar\sigma}\rangle X^{\bar\sigma 2} ,
\end{eqnarray}
and $A_{pq}= \langle X^{pp}+X^{qq}\rangle $;
$A_{0\sigma}=1-n_{\bar\sigma}$, $A_{2 \bar\sigma}=n_{\bar\sigma}$.

In this case
\begin{eqnarray}
 \alpha_1^{0\sigma}& =& -\zeta\alpha_1^{\bar\sigma 2} =
 \frac{V}{A_{0\sigma}} \varphi_\sigma, \nonumber \\
 \alpha_2^{0\sigma} &=& -\zeta\alpha_2^{\bar\sigma 2} =
 -\frac{V}{A_{2 \bar\sigma}} \zeta \varphi_\sigma, \nonumber \\
 \varphi_\sigma &=&
 \langle \xi_{\bar\sigma} X^{\bar\sigma 0}\rangle +
 \zeta \langle X^{\sigma 2} \xi_{\bar\sigma}^{+} \rangle.
\end{eqnarray}

Using this procedure by differentiating both with respect to the
left and to the right time arguments, we come to the relations
between the components of the Green's function $G_\sigma^{(a)}$
and scattering matrix $\hat{P_\sigma}$. In a matrix
representation, we have
\begin{equation}
 \hat{G_\sigma}=\hat{G_0^\sigma}+\hat{G_0^\sigma}\hat{P_\sigma}\hat{G_0^\sigma},
\end{equation}
where
\begin{equation}
  \hat{G_\sigma}=2\pi \left(
  \begin{array}{lll}
  \langle\langle X^{0\sigma}| X^{\sigma 0} \rangle\rangle & &
  \langle\langle X^{0\sigma}| X^{2 \bar\sigma} \rangle\rangle \\
  \langle\langle X^{\bar\sigma 2}| X^{\sigma 0} \rangle\rangle & &
  \langle\langle X^{\bar\sigma 2}| X^{2 \bar\sigma} \rangle\rangle
  \end{array}
  \right),
  \label{Grin_Mart}
\end{equation}
and nonperturbed Green's function $\hat {G^\sigma_0}$ is
\begin{equation}
  \hat{G}^\sigma_0=\frac{1}{D_\sigma} \left(
  \begin{array}{lll}
  \omega-b_\sigma & & -\zeta\frac{V}{A_{2\bar\sigma}}\varphi_\sigma \\
  -\zeta\frac{V}{A_{0\sigma}}\varphi_\sigma & & \omega-a_\sigma
  \end{array}
  \right) \left(
  \begin{array}{lll}
  A_{0\sigma} & & 0 \\
  0  & & A_{2\bar\sigma}
  \end{array}
  \right),
\end{equation}
where
\begin{equation}
  D_\sigma=(\omega-a_\sigma)(\omega-b_\sigma)-
  \frac{V^2}{A_{0\sigma}A_{2\bar\sigma}}\varphi_\sigma^2,
\end{equation}
\begin{equation}
  a_\sigma=-\mu_\sigma+\frac{V}{A_{0\sigma}}\varphi_\sigma,
  \qquad
  b_\sigma=U-\mu_\sigma+\frac{V}{A_{2\bar\sigma}}\varphi_\sigma.
\end{equation}
The scattering matrix
\begin{equation}
  \hat{P}_\sigma=2\pi
  \left(
  \begin{array}{lll}
  A_{0\sigma}^{-1} & & 0 \\
  0  & & A_{2\bar\sigma}^{-1}
  \end{array}
  \right)
  \left(
  \begin{array}{lll}
  \langle\langle Z^{0\sigma}| Z^{\sigma 0} \rangle\rangle & &
  \langle\langle Z^{0\sigma}| Z^{2 \bar\sigma} \rangle\rangle \\
  \langle\langle Z^{\bar\sigma 2}| Z^{\sigma 0} \rangle\rangle & &
  \langle\langle Z^{\bar\sigma 2}| Z^{2 \bar\sigma} \rangle\rangle
  \end{array}
  \right)
  \left(
  \begin{array}{lll}
  A_{0\sigma}^{-1} & & 0 \\
  0  & & A_{2\bar\sigma}^{-1}
  \end{array}
  \right)
  \label{scattmatr}
\end{equation}
being expressed in terms of irreducible Green's functions contains
the scattering corrections of the second and the higher orders in
powers of $V$. The separation of the irreducible parts in
$\hat{P_\sigma}$ enables us to obtain the mass operator
$\hat{M_\sigma}$ and the single-site Green's function expressed as
a solution of the Dyson equation
\begin{equation}
 \hat{G_\sigma}=(1-\hat{G^\sigma_0}\hat{M}_\sigma)^{-1} \hat{G^\sigma_0}.
 \label{dyson1}
\end{equation}

We will restrict ourselves hereafter to the simple approximation
in calculating the mass operator $\hat {P_\sigma}$, taking into
account the scattering processes of the second order in $V$. In
this case
\begin{equation}
 \hat M_{\sigma}= \hat P_{\sigma}^{(0)},
 \label{massop}
\end{equation}
where the irreducible Green's functions are calculated without
allowance for correlation between electron transition on the given
site and environment. It corresponds to the procedure of
different-time decoupling \cite{decoup}, which means in our case
an independent averaging of the products of $X$ and $\xi$
operators. Let us illustrate this approximation by some examples.
\begin{enumerate}  
\item The Green's function $ \langle\langle
\overline{(X^{00}+X^{\sigma\sigma})\xi_\sigma^{\phantom{+}}} |
\overline{\xi_\sigma^{+}(X^{00}+X^{\sigma\sigma})}
\rangle\rangle_\omega \equiv I_1 (\omega)$.

Using the spectral theorem and performing the different-time
decoupling:
\begin{equation}
 \langle
  \xi_\sigma^{+}(t)(X^{00}+X^{\sigma\sigma})_t
  (X^{00}+X^{\sigma\sigma})\xi_\sigma
 \rangle^{\mathrm{ir}} \approx
 \langle (X^{00}+X^{\sigma\sigma})_t
 (X^{00}+X^{\sigma\sigma}) \rangle
 \langle
  \xi_\sigma^{+}(t) \xi_\sigma
 \rangle,
 \label{e43}
\end{equation}
we obtain
\begin{equation}
  I_1(\omega)=A_{0\sigma}\langle\langle \xi_\sigma^{\phantom{+}}| \xi_\sigma^{+}
  \rangle\rangle_\omega =
  \frac{A_{0\sigma}}{2\pi V^2} J_\sigma (\omega).
\end{equation}
Here, $X$-correlators are in a zero approximation
\begin{equation}
 \langle (X^{00}+X^{\sigma\sigma})_t
 (X^{00}+X^{\sigma\sigma}) \rangle \approx
 \langle (X^{00}+X^{\sigma\sigma})^2 \rangle = A_{0\sigma}\,.
\end{equation}

\item The Green's function $ \langle\langle
\overline{X^{\bar\sigma\sigma}\xi_{\bar\sigma}^{\phantom{+}}} |
\overline{\xi_{\bar\sigma}^{+} X^{\sigma\bar\sigma}}
\rangle\rangle_\omega \equiv I_2 (\omega)$.

The time correlation function $ \langle
  \xi_{\bar\sigma}^{+}(t)X^{\sigma \bar\sigma}(t)
  X^{\bar\sigma \sigma}\xi_{\bar\sigma}^{\phantom{+}}
 \rangle^{\mathrm{ir}}$ is decoupled as
\begin{equation}
 \langle
  \xi_{\bar\sigma}^{+}(t)X^{\sigma \bar\sigma}(t)
  X^{\bar\sigma \sigma}\xi_{\bar\sigma}^{\phantom{+}}
 \rangle^{\mathrm{ir}}  \approx
 \langle X^{\sigma \bar\sigma}(t) X^{\bar\sigma \sigma} \rangle
 \langle \xi_{\bar\sigma}^{+}(t) \xi_{\bar\sigma}^{\phantom{+}} \rangle.
\end{equation}
In the zero approximation
\begin{equation}
  \langle X^{\sigma \bar\sigma}(t) X^{\bar\sigma \sigma} \rangle=
  \exp \big[ \ri (\mu_{\bar\sigma}-\mu_\sigma)t\big]
 \langle X^{\sigma\sigma} \rangle.
\end{equation}
Using these expressions we obtain
\begin{eqnarray}
2\pi V^2 I_2(\omega)&=&\frac{\langle
X^{\sigma\sigma}\rangle+
  \langle X^{\bar\sigma\bar\sigma}\rangle}{2}
  J_{\bar\sigma}(\omega+\mu_\sigma-\mu_{\bar\sigma})
  -\frac{\langle X^{\sigma\sigma}\rangle
  -\langle X^{\bar\sigma\bar\sigma}\rangle}{4\pi}
  \nonumber \\
&& \times \int_{-\infty}^{+\infty} \frac{\rd
\omega'}{\omega-\omega'-\mu_{\bar\sigma}+\mu_\sigma}
  \tanh\frac{\beta\omega'}{2}\big\{-2\Imm
  J_{\bar\sigma}(\omega'+\ri \varepsilon)\big\}.
\qquad\mathstrut
\end{eqnarray}
Let us notice that in the case with $n_A=n_B$, $\mu_A=\mu_B$,
which corresponds to the simple Hubbard model in the absence of an
external magnetic field,
\begin{equation}
  I_2(\omega) = \frac{\langle X^{\sigma\sigma}\rangle}{2\pi V^2}
  J_{\bar\sigma}(\omega).
\end{equation}
\item The Green's function $ \langle\langle
\overline{X^{02}\xi_{\bar\sigma}^{+}} |
\overline{\xi_{\bar\sigma}^{\phantom{+}} X^{20}}
\rangle\rangle_\omega \equiv I_3 (\omega)$.

In this case
\begin{eqnarray}
I_3(\omega)  &=&  \frac{\langle X^{00}\rangle+
  \langle X^{22}\rangle}{2}
  \langle\langle
   \xi_{\bar\sigma}^{+} | \xi_{\bar\sigma}^{\phantom{+}}
  \rangle\rangle_{\omega+\mu_\sigma+\mu_{\bar\sigma}-U}
 {}+\frac{\langle X^{00}\rangle-\langle X^{22}\rangle}{4\pi}
 \nonumber \\
&& \times  \int_{-\infty}^{+\infty} \frac{\rd
\omega'}{\omega-\omega'+\mu_{\bar\sigma}+\mu_\sigma-U}
  \tanh\frac{\beta\omega'}{2}\big\{-2\Imm
    \langle\langle
   \xi_{\bar\sigma}^{+} | \xi_{\bar\sigma}^{\phantom{+}}
  \rangle\rangle_{\omega'+\ri \varepsilon} \big\}.
\qquad
\end{eqnarray}

Using the above results, the mass operator can be obtained in an
explicit form. According to (\ref{massop}), it is given by the
expression (\ref{scattmatr}) where
\begin{eqnarray}
  \langle\langle Z^{0\sigma}|Z^{\sigma 0}\rangle\rangle_\omega &=&
    A_{0\sigma} J_\sigma (\omega)-R_\sigma (\omega), \nonumber \\
  \langle\langle Z^{\bar\sigma 2}|Z^{2\bar\sigma}\rangle\rangle_\omega &=&
    A_{2\bar\sigma} J_\sigma (\omega) -R_\sigma (\omega), \nonumber \\
  \langle\langle Z^{0\sigma}|Z^{2\bar\sigma}\rangle\rangle_\omega &=&
  \langle\langle Z^{\bar\sigma 2}|Z^{\sigma
  0}\rangle\rangle_\omega= R_\sigma (\omega),
  \label{zzzz}
\end{eqnarray}
and
\begin{eqnarray}
  \lefteqn {R_\sigma(\omega)  =  -\frac{\langle X^{\sigma\sigma}\rangle+
  \langle X^{\bar\sigma\bar\sigma}\rangle}{2}
  J_{\bar\sigma}(\omega+\mu_\sigma-\mu_{\bar\sigma})}  \nonumber\\
  & & {}+\frac{\langle X^{\sigma\sigma}\rangle-\langle X^{\bar\sigma\bar\sigma}\rangle}{4\pi}
  \int_{-\infty}^{+\infty}\! \frac{\rd \omega'}{\omega-\omega'-\mu_{\bar\sigma}+\mu_\sigma}
  \big\{{-}2\Imm
  J_{\bar\sigma}(\omega'+\ri\varepsilon)\big\}\tanh\frac{\beta\omega'}{2} \nonumber\\
  & &{{}+\frac{\langle X^{0 0}\rangle+
  \langle X^{2 2}\rangle}{2}
  J_{\bar\sigma}(U-\mu_\sigma-\mu_{\bar\sigma}-\omega)} \nonumber \\
  & & {}+\frac{\langle X^{00}\rangle-\langle X^{22}\rangle}{2\pi}
  \int_{-\infty}^{+\infty}\! \frac{\rd \omega'}{\omega-\omega'+\mu_{\bar\sigma}+\mu_\sigma-U}
  \big\{{-}\Imm  J_{\bar\sigma}(-\omega'-\ri \varepsilon)\big\}
  \tanh\frac{\beta\omega'}{2}.
\nonumber \\
\end{eqnarray}
\end{enumerate}

\section{\mathversion{bold}\mathsurround=0pt%
Set of DMFT equations in the infinite-$U$ limit}

The following consideration will be performed in the case of $U
\rightarrow +\infty$ which excludes simultaneous occupation by two
particles of $A$ and $B$ types of the same site, when the model is
used in describing the lattice gas of the particles of two types.
Then functions $R_\sigma$ for $\sigma=A$ or $B$ are as follows
\arraycolsep=2pt
\begin{eqnarray}
  R_A(\omega) & =&  -\frac{n_A+n_B}{2}
  J_B(\omega+\mu_A-\mu_B)    \nonumber \\
  & & {}+\frac{n_A-n_B}{4\pi}
  \int_{-\infty}^{+\infty} \frac{\rd \omega'}{\omega-\omega'-\mu_B+\mu_A}
  \tanh\frac{\beta\omega'}{2}\big\{-2\Imm
  J_B(\omega'+\ri \varepsilon)\big\},\qquad
  \label{ss3}
\end{eqnarray}
\begin{eqnarray}
  R_B(\omega) & =&  -\frac{n_A+n_B}{2}
  J_A(\omega+\mu_B-\mu_A)  \nonumber \\
  & & {}+\frac{n_B-n_A}{4\pi}
  \int_{-\infty}^{+\infty} \frac{\rd \omega'}{\omega-\omega'-\mu_A+\mu_B}
  \tanh\frac{\beta\omega'}{2}\big\{-2\Imm
  J_A(\omega'+\ri \varepsilon)\big\},\qquad
  \label{ss4}
\end{eqnarray}
The single-site Green's functions (\ref{Grin_X}) are obtained
using relations (\ref{dyson1}), (\ref{massop}) and (\ref{zzzz}) in
the $U\rightarrow+\infty$ limit, and respectively they equal to:
\begin{equation}
  G^{(a)}_A (\omega)=\frac{1-n_B}{\omega+\mu_A-\frac{V}{1-n_B}\varphi_A-J_A(\omega)+\frac{R_A(\omega)}{1-n_B}},
  \label{ss1}
\end{equation}
\begin{equation}
  G^{(a)}_B (\omega)=\frac{1-n_A}{\omega+\mu_B-\frac{V}{1-n_A}
  \varphi_B-J_B(\omega)+\frac{R_B(\omega)}{1-n_A}}.
  \label{ss2}
\end{equation}

Parameter $\varphi_\sigma$, which is expressed by the average
values of the products of $X$ and $\xi$ operators, is a functional
of the potential $J_\sigma(\omega)$. According to the spectral
theorem
\begin{eqnarray}
   V\langle X^{\sigma 0}\xi_\sigma \rangle &=&\ri \int_{-\infty}^{+\infty}
   \frac{\rd \omega}{\re ^{\beta\omega}+1}\big[
V\langle\langle \xi_\sigma|X^{\sigma 0}\rangle\rangle_{\omega+\ri
\varepsilon}- V\langle\langle \xi_\sigma|X^{\sigma
0}\rangle\rangle_{\omega-\ri \varepsilon}
   \big].
\end{eqnarray}
The Green's functions are found using linearized equations of
motion and neglecting the irreducible parts.

In this case
\begin{equation}
  V\varphi_A=-\frac{1}{2\pi}\int_{-\infty}^{+\infty}
   \frac{\rd \omega}{\re ^{\beta \omega}+1}
   \bigg\{ -2\Imm \frac{(1-n_A) J_B(\omega)}{\omega+\mu_B-V\frac{\varphi_B}{1-n_A}}
   \bigg\}_{\omega+\ri \varepsilon},
   \label{ss5}
\end{equation}
\begin{equation}
  V\varphi_B=-\frac{1}{2\pi}\int_{-\infty}^{+\infty}
   \frac{\rd \omega}{\re ^{\beta \omega}+1}
   \bigg\{ -2\Imm \frac{(1-n_B) J_A(\omega)}{\omega+\mu_A-V\frac{\varphi_A}{1-n_B}}
   \bigg\}_{\omega+\ri \varepsilon}.
   \label{ss6}
\end{equation}

The coherent potential $J_{\sigma}(\omega)$ is self-consistently
determined from the equations (\ref{sys2})--(\ref{sys3}) by
eliminating the total irreducible part. Integration with the
semielliptic
\begin{equation}
  \rho_\sigma(\varepsilon)=\frac{2}{\pi W_\sigma^2} \sqrt{W_\sigma^2-\varepsilon^2}
\end{equation}
density of states is done. In this case, we have simple
expressions
\begin{equation}
  G^{(a)}_A(\omega)=\frac{4J_A(\omega)}{W_A^2},
  \label{ss7}
\end{equation}
\begin{equation}
  G^{(a)}_B(\omega)=\frac{4J_B(\omega)}{W_B^2}.
  \label{ss8}
\end{equation}
The set of equations becomes a closed one by adding expressions
for average particle concentrations obtained with the help of the
imaginary parts of the Green's functions (i.e. interacting
densities of states):
\begin{equation}
  n_A=\frac{1}{2\pi}\int_{-\infty}^{+\infty}\rd \omega \frac{-2\Imm
  G^{(a)}_A(\omega+\ri \varepsilon)}{\re ^{\beta\omega}+1},
  \label{ss9}
\end{equation}
\begin{equation}
  n_B=\frac{1}{2\pi}\int_{-\infty}^{+\infty}\rd \omega \frac{-2\Imm
  G^{(a)}_B(\omega+\ri \varepsilon)}{\re ^{\beta\omega}+1}.
  \label{ss10}
\end{equation}

It has been shown in the case of the standard Hubbard model
\cite{ista1} that such a set of equations corresponds to GH3
approximation, and includes, as some special cases, a number of
known approximations. The case with $R_\sigma=0$,
$\varphi_\sigma=0$ corresponds to the alloy-analogy (AA)
approximation \cite{alloya1}. The modified alloy-analogy (MAA)
approximation is obtained by taking into account the
renormalization of the local electron levels $\varphi_\sigma$
\cite{alloya1,alloya2}. For $R_\sigma\neq 0$, $\varphi_\sigma=0$
the system of equations corresponds to the extension of the
Hubbard-III approximation by the inclusion of the integral terms
responsible for scattering.

\section{Lattice gas thermodynamics in the Falicov-Kimball limit}
\subsection{Exact solution of DMFT problem}

In the limiting case of $U \rightarrow +\infty$, $W_B \rightarrow
0$ ($t_{ij}^B \rightarrow 0$), when mobility of the $B$ type ions
is very small (this case corresponds to the spinless
Falicov-Kimball model \cite{FK1} with the infinite on-site
repulsion), the effective single-site problem can be solved
exactly. Thermodynamics of the model, equilibrium states and phase
transitions were investigated in a series of papers
(\cite{Freericks1,Falicov1,Falicov2}, see \cite{Fr_rev} as well).
There were considered thermodynamic regimes with the constant
 $n=n_A+n_B$, $\varepsilon = \mu_B-\mu_A$ values \cite{Falicov1}
and with the fixed relative filling $n_B=\mathrm{const}$, $n_A = \mathrm{const}
\cdot (1-n_B) $ \cite{Freericks1}. However, other thermodynamic
regimes should be also considered: (i) constant chemical
potentials $\mu_A$, $\mu_B$; (ii) the mixed regimes $\mu_A=\mathrm{const}$,
$n_B=\mathrm{const}$ or $\mu_B=\mathrm{const}$, $n_A=\mathrm{const}$. Let us note that in the
case of the pseudospin-electron model (PEM) without tunnelling,
the dynamics of pseudospins being analogous to the Falicov-Kimball
model, the regime of $\mu_B=\mathrm{const}$ corresponds to the fixation of
longitudinal asymmetry field $h$ acting on pseudospins. The PEM
thermodynamics has been a subject of investigations within DMFT
\cite{PEM} as well as in the framework
of approximate methods
(generalized random phase approximation (GRPA),
\cite{GRPA,GRPA2,GRPA3}).

An exact solution of the single-site problem is given by the
simple expression:
\begin{equation}
  G^{(a)}_A (\omega)=\frac{1-n_B}{\omega+\mu_A-J_A (\omega)}.
  \label{GaFK}
\end{equation}
It should be mentioned that this formula can be obtained by the
above described approach and corresponds to the AA approximation.
The final expression for the Green's function is obtained by
solving the quadratic equation formed by (\ref{ss7}),
(\ref{GaFK}):
{\arraycolsep=1pt
\begin{eqnarray}
  \lefteqn{G^{(a)}_A (\omega \pm \ri \varepsilon)=} \nonumber \\
&&=
  \left\{
  \begin{array}{lll}
     \frac{2}{W_A^2}(\omega+\mu_A) + \frac{2}{W_A^2} \sqrt{(\omega+\mu_A)^2-W_A^2(1-n_B)}, && \omega+\mu_A<-W_A\sqrt{1-n_B}; \\ [1.2ex]
     \frac{2}{W_A^2}(\omega+\mu_A) \mp \ri \frac{2}{W_A^2} \sqrt{W_A^2(1-n_B)-(\omega+\mu_A)^2}, &&|\omega+\mu_A|\leqslant W_A\sqrt{1-n_B}; \\ [1.2ex]
     \frac{2}{W_A^2}(\omega+\mu_A) - \frac{2}{W_A^2} \sqrt{(\omega+\mu_A)^2-W_A^2(1-n_B)}, && \omega+\mu_A>W_A\sqrt{1-n_B}.
  \end{array}
  \right.\qquad
  \label{GFK}
\end{eqnarray}
}%
Here, the phase of the square root is chosen so that an imaginary
part of the Green's function has the correct sign as well as by
using the properties which follow from the spectral theorem.

Using this Green's function the first equation for the particle
concentrations is obtained from (\ref{ss9})
\begin{equation}
  n_A=\frac{2(1-n_B)}{\pi} \int_{-1}^{+1} \rd x \frac{\sqrt{1-x^2}}{\re ^{\beta
  (W_A\sqrt{1-n_B}
  x-\mu_A)}+1}.
  \label{FFF12}
\end{equation}
Here, in the standard scheme, the second equation is obtained
thermodynamically by differentiating the grand canonical potential
\cite{ista2}
\begin{equation}
  n_B=\frac{\re ^{\beta\mu_B+Q}}{1+\re ^{\beta\mu_A}+\re ^{\beta\mu_B+Q}},
  \label{FFF1}
\end{equation}
where in our case
\begin{equation}
Q=-\beta \frac{W_A \sqrt{1-n_B}}{\pi} \int_{-1}^{1}\frac{\rd
x}{\re ^{-\beta(xW_A \sqrt{1-n_B}-\mu_A))}+1}
\arctan\frac{\sqrt{1-x^2}}{x}. \label{FFF123}
\end{equation}

Let us notice that parameter $Q$ does not depend on the sign of
chemical potential of the moving particles, i.e., it does not
depend on the interchange $\mu_A\leftrightarrow -\mu_A$.

Separating the solutions of a set of equations
(\ref{FFF12})--(\ref{FFF123}) that correspond to absolute minima
of the grand canonical potential, we will investigate equilibrium
states in the above-mentioned thermodynamic regimes. There are
phase transitions between homogeneous phases with different
particle concentrations in the regime of constant chemical
potentials ($\mu_A=\mathrm{const}$, $\mu_B=\mathrm{const}$). Phase separation
phenomena take place in the regimes where either of the particle
concentrations is constant.

\begin{figure}
\includegraphics[width=0.45\textwidth]{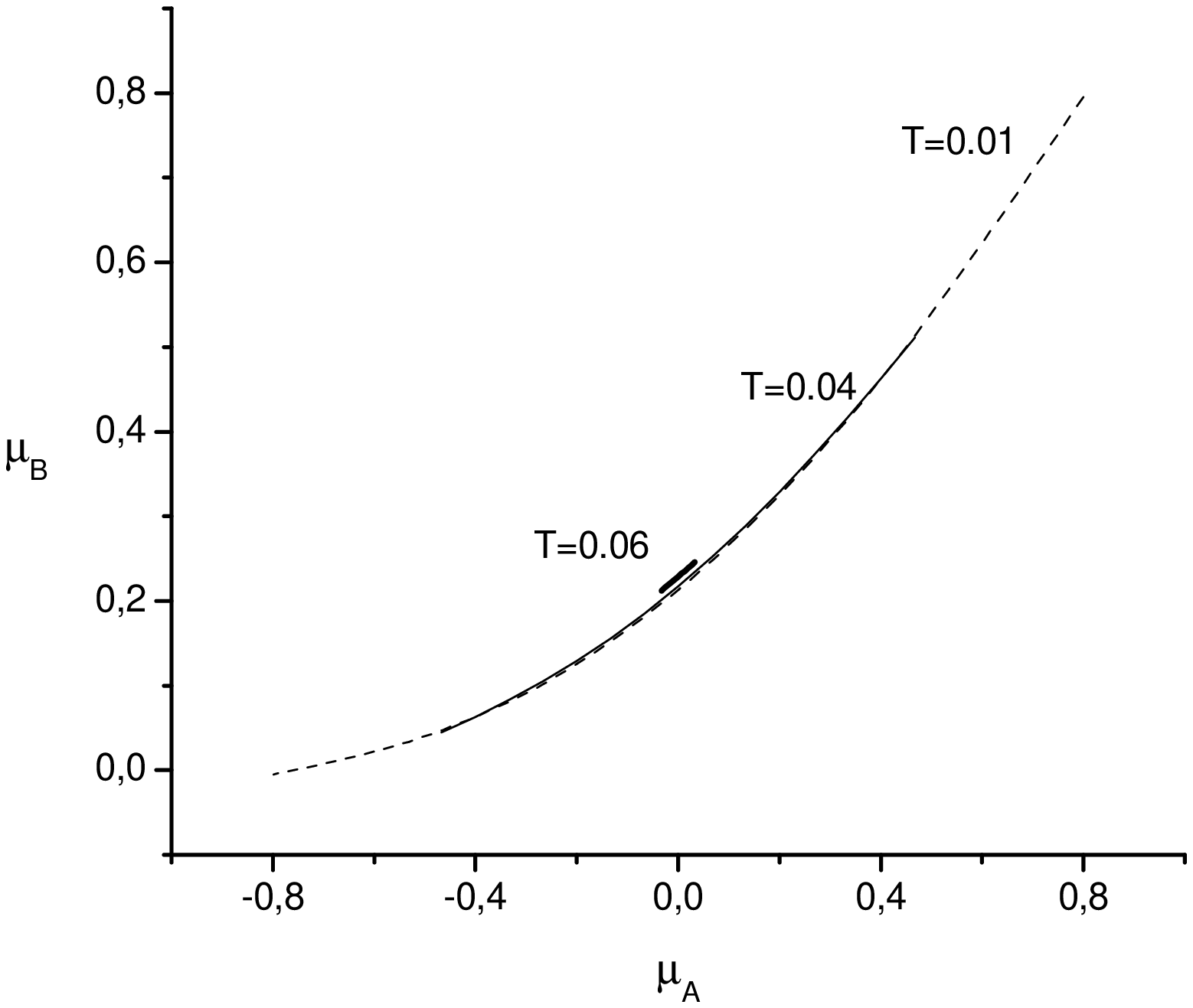}
\hfill
\includegraphics[width=0.45\textwidth]{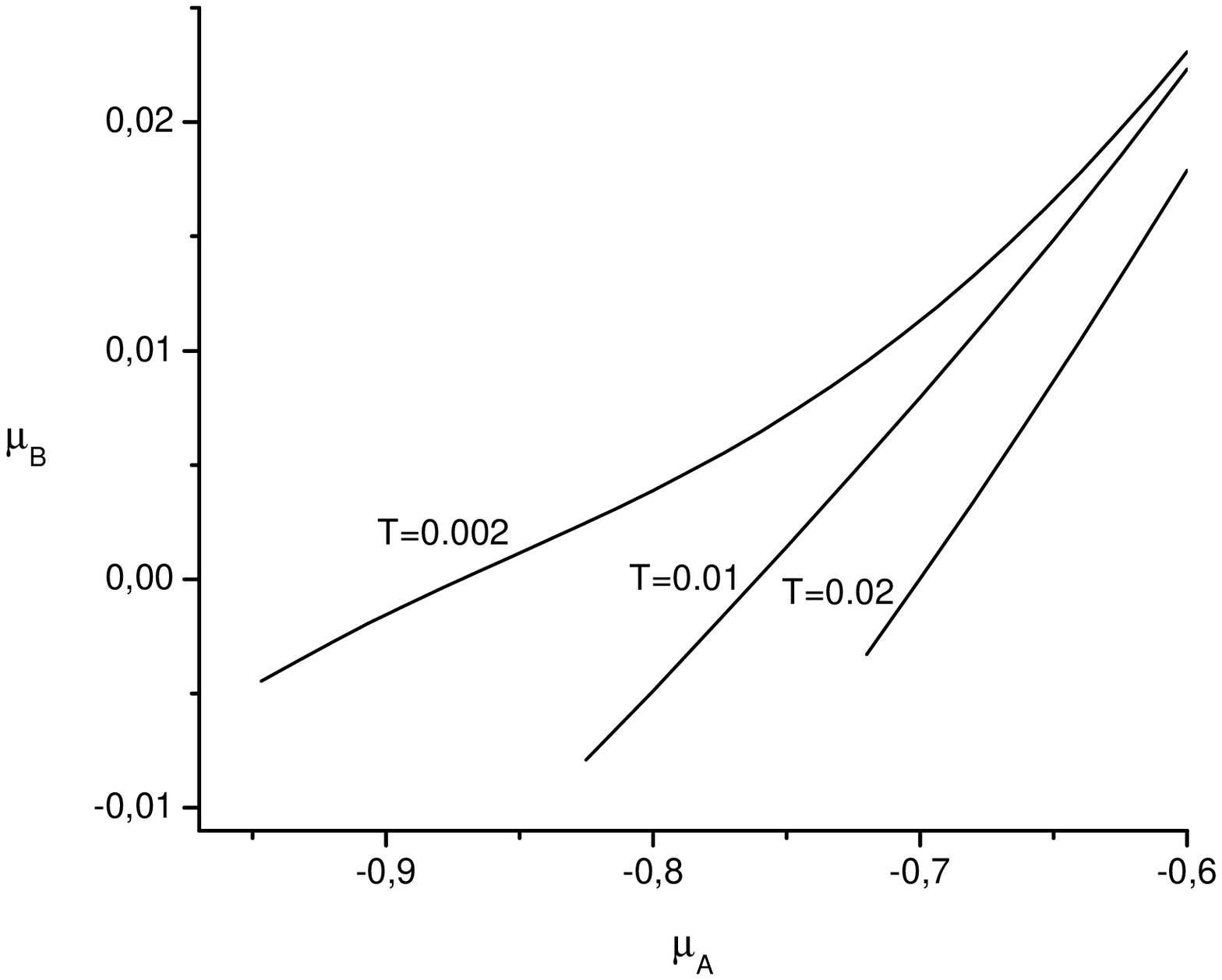}
\\
\parbox[t]{0.49\textwidth}{
\caption{$(\mu_A,\mu_B)$-phase diagrams. The parameter values
$W_A=1$, $W_B=0$, the dashed line -- $T=0.01$, the solid line --
$T=0.04$, the bold line -- $T=0.06$. } \label{Fig_mm}} \hfill
\parbox[t]{0.49\textwidth}{\caption{The part of $(\mu_A,\mu_B)$-phase
 diagrams at different temperatures; the appearance of the phase
transitions for $\mu_B<0$ ($W_A=1$, $W_B=0$) is shown.}
\label{Fig_mm2}}
\end{figure}

\begin{enumerate}%
\item $\mu_A=\mathrm{const}$, $\mu_B=\mathrm{const}$. In figure~\ref{Fig_mm}
phase diagrams are shown in ($\mu_A, \mu_B$) coordinates. All
energy quantities are given in the units of a half-bandwidth
($W_A=1$). The phase transition curves terminate at critical points.
They become shorter (symmetrically with respect to the value
$\mu_A=0$) with the temperature growth and vanish at critical
temperature. Numerical calculations give the following values of
critical parameters
\begin{equation}
 T^{\mathrm{c}} \approx 0.0601 W_A, \qquad
 \mu_A^{\mathrm{c}}  =  0,  \qquad
 \mu_B^{\mathrm{c}}  \approx  0.229 W_A.
\label{crit}
\end{equation}
At the zero temperature, phase transitions are within the
$-W_A<\mu_A<W_A$, $0<\mu_B<W_A$ ranges of chemical potentials.
However, figure~\ref{Fig_mm2} shows that there is  a possibility
of phase transitions with $\mu_B<0$ when temperature increases.

Let us note that diagrams analogous to those in
figure~\ref{Fig_mm} can be obtained in DMFT for PEM (where $\mu_A$
corresponds to chemical potential $\mu$ of electrons, and $\mu_B$
corresponds to the asymmetry field $h$). The phase coexistence
curve on $(\mu,h)$ diagram obtained for finite values of the
coupling constant in \cite{PEM} consists of two parts
corresponding to the chemical potential being in the vicinity or
inside of either electron subband (the band splitting in a
spectrum is caused by interaction with pseudospins). There is a
direct correspondence between the lower part of the phase diagram
for PEM and the diagrams in figure~\ref{Fig_mm2}.

\item $\mu_A=\mathrm{const}$, $n_B=\mathrm{const}$ (figure~\ref{Fig_Tnb}(a)). In
this case, a phase separation takes place. A topology of the phase
diagrams ($n_B,T$) depends only on the absolute value of the
chemical potential of the moving particles. The highest
value of critical temperature is reached
at $\mu_A=0$ and is given by the expression
(\ref{crit}). The critical temperature decreases when chemical
potential of the moving particles approaches either of the band
edges. At the zero temperature for $-W_A<\mu_A<W_A$, the system is
separated into two phases: $n_B^{(1)}=0$ and $n_B^{(2)}=1$. The
regime of relative half-filling ($n_B=\mathrm{const}$, $n_A=(1-n_B)/2$) has
been considered in \cite{Freericks1}. This case corresponds to the
$n_B=\mathrm{const}$ regime with $\mu_A=0$, so the diagram which is
equivalent to figure~\ref{Fig_Tnb}(a) for $\mu_A=0$ has been
previously obtained.

\begin{figure}
\includegraphics[width=0.32\textwidth]{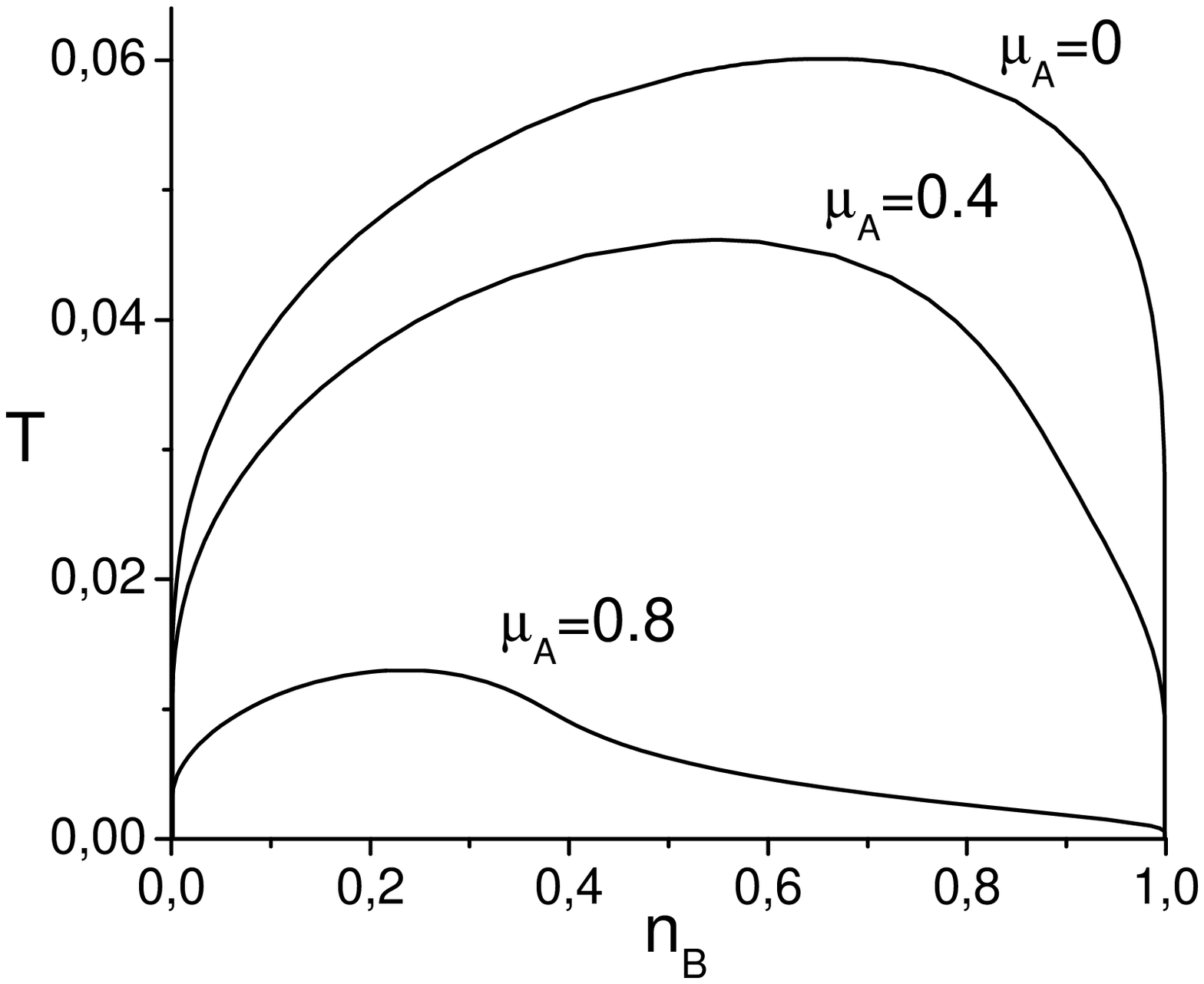}
\hfill
\includegraphics[width=0.32\textwidth]{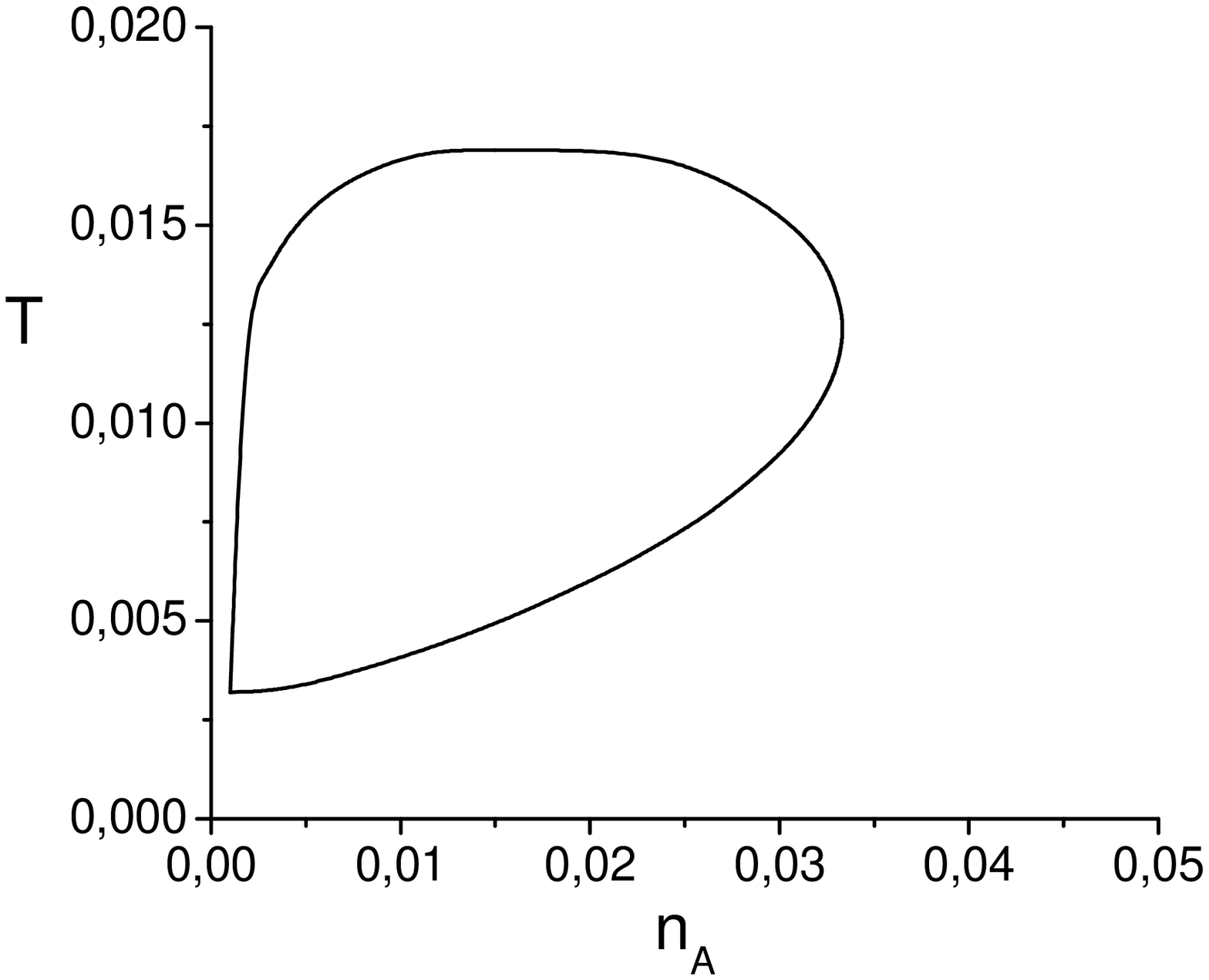}
\hfill
\includegraphics[width=0.32\textwidth]{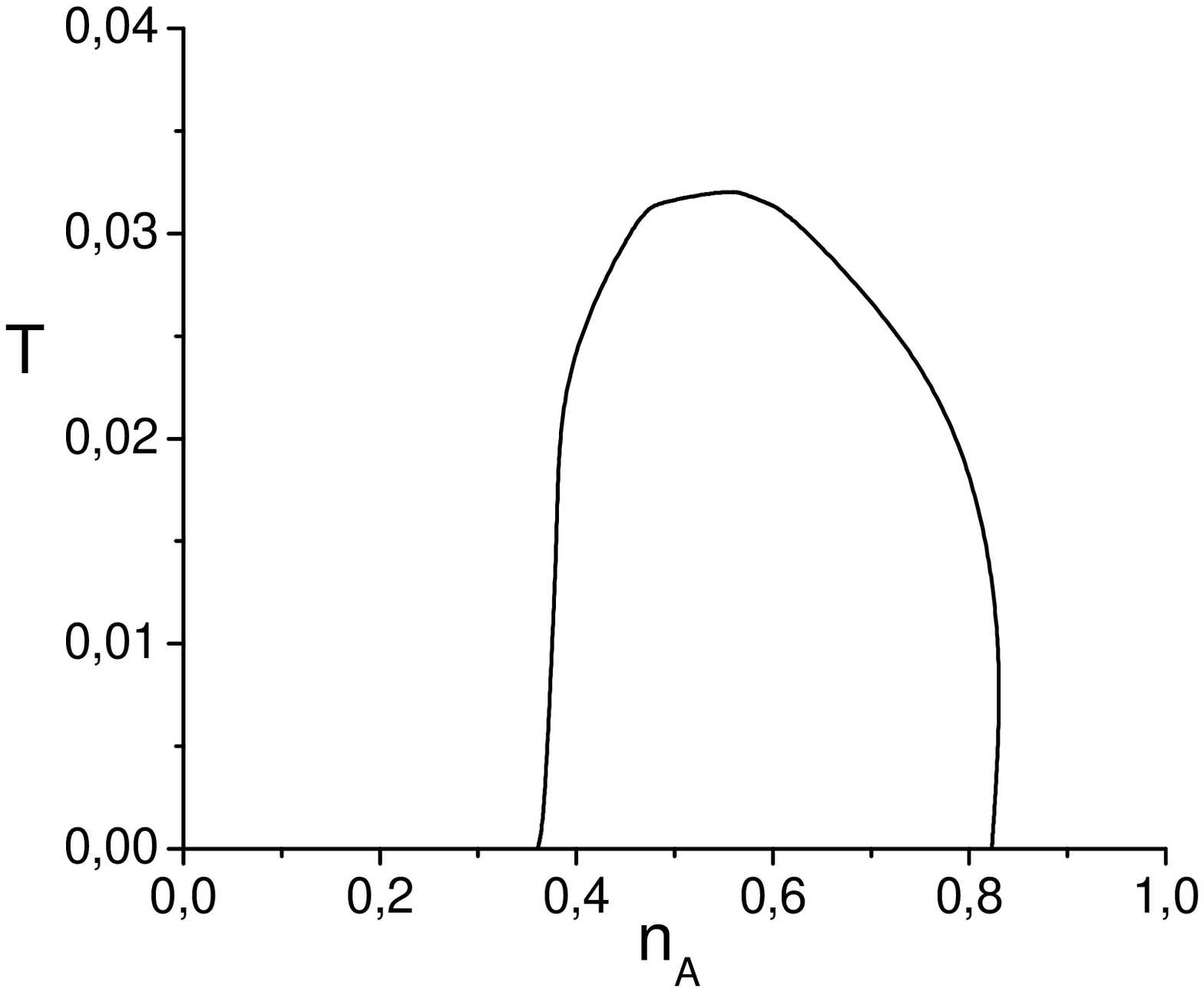}
\\
\parbox[t]{0.32\textwidth}{\centerline{(a)}}
\hfill
\parbox[t]{0.32\textwidth}{\centerline{(b)}}
\hfill
\parbox[t]{0.32\textwidth}{\centerline{(c)}}
\caption{The phase separation diagrams for $W_A=1$, $W_B=0$. (a)
-- $(n_B, T)$-phase diagrams for the different chemical potentials
of moving particles. (b), (c) -- $(n_A, T)$-phase diagrams for
$\mu_B=-0.006$ and $\mu_B=0.6$ respectively.} \label{Fig_Tnb}
\end{figure}

\item $n_A=\mathrm{const}$, $\mu_B=\mathrm{const}$ (figure~\ref{Fig_Tnb}(b,c)).
There is a homogeneous state at zero temperature, but a segregated
state appears at higher temperatures for $-0.009W_A<\mu_B<0$
(figure~\ref{Fig_Tnb}(b)). The phase separation region decreases
continuously and there is a point at temperature $T\approx
0.01~W_A$ when chemical potential of localized particles
approaches some critical value $\mu_B \approx -0.009~W_A$. The
possibility of such a situation is illustrated by the behaviour of
the left end of the phase transition curve in the vicinity of
$\mu_B=0$, $\mu_A=-W_A$ values (figure~\ref{Fig_mm2}). The
thermodynamic regime with $\varepsilon^0=\mu_A-\mu_B=\mathrm{const}$,
$n_A+n_B=\mathrm{const}$ has been investigated in \cite{Falicov1}, and
phase diagrams similar to that in figure~\ref{Fig_Tnb}(b) have
been obtained. It can be seen that both ends of the phase
transition curve ($\mu_A, \mu_B$) have the similar temperature
behaviour (with respect to $\mu_B=0$ and to $\mu_B-\mu_A=0$
lines).

At $\mu_B>0$, the topological structure of phase diagram $(n_A,T)$
is similar to the one in figure~\ref{Fig_Tnb}(c). A phase
separated state always exists for low temperatures with
non-integer values of $n_A$ in both phases at $T=0$. Diagrams of
this type have been obtained for PEM in the strong coupling
case (that corresponds to $U\gg W_A$) using a GRPA approach
\cite{GRPA2}.
\end{enumerate}%

\begin{figure}
\includegraphics[width=0.32\textwidth]{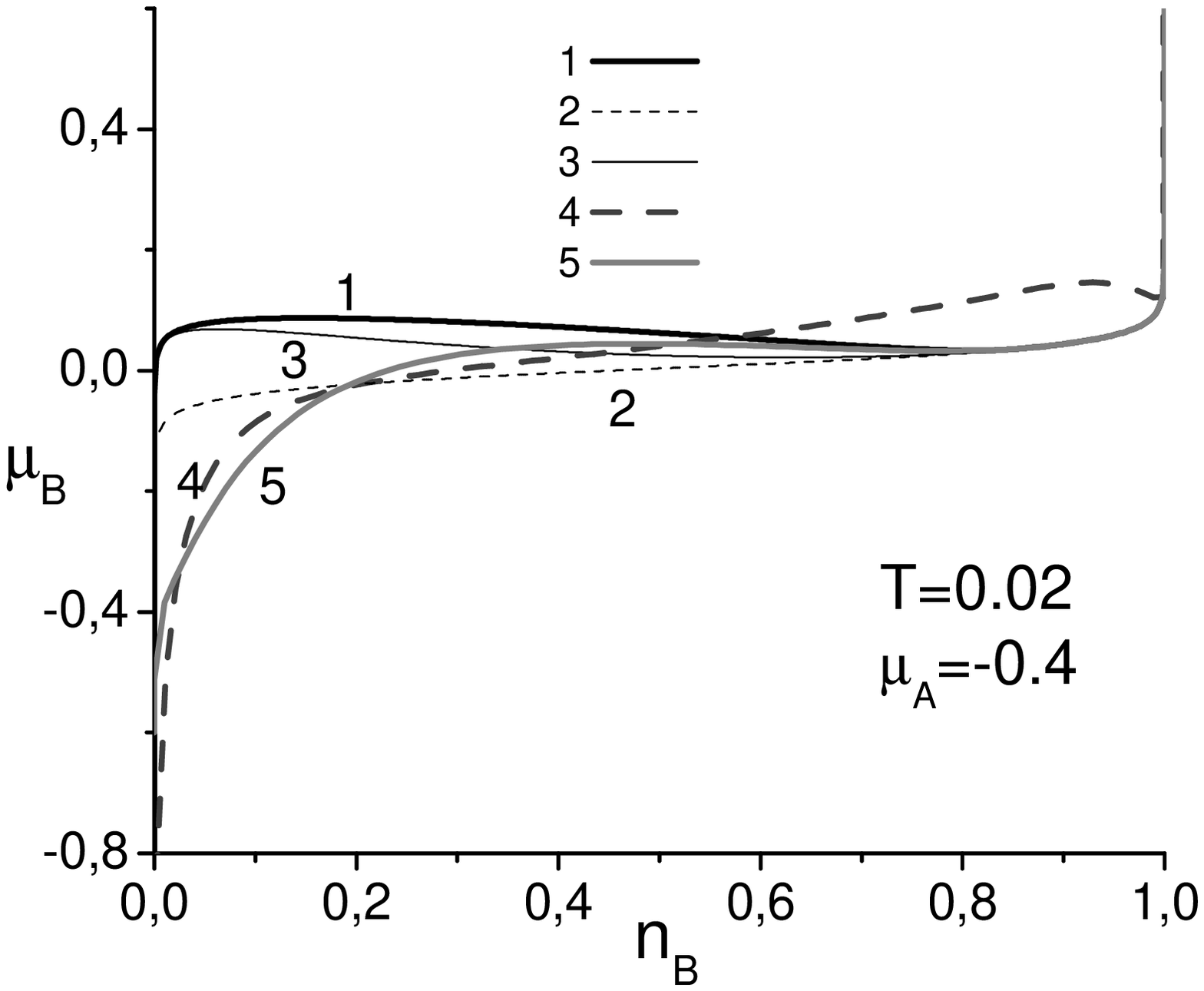}
\hfill
\includegraphics[width=0.32\textwidth]{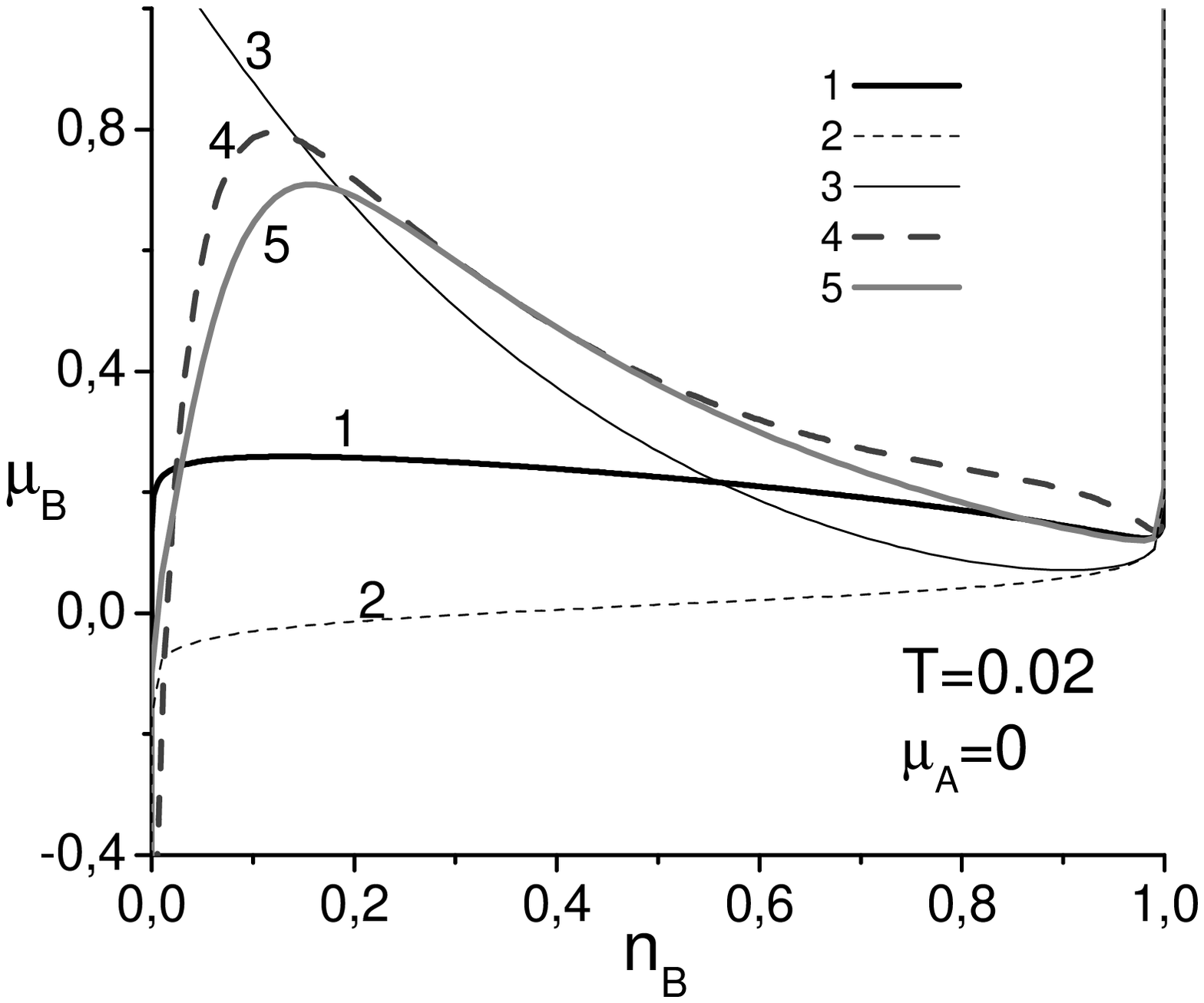}
\hfill
\includegraphics[width=0.32\textwidth]{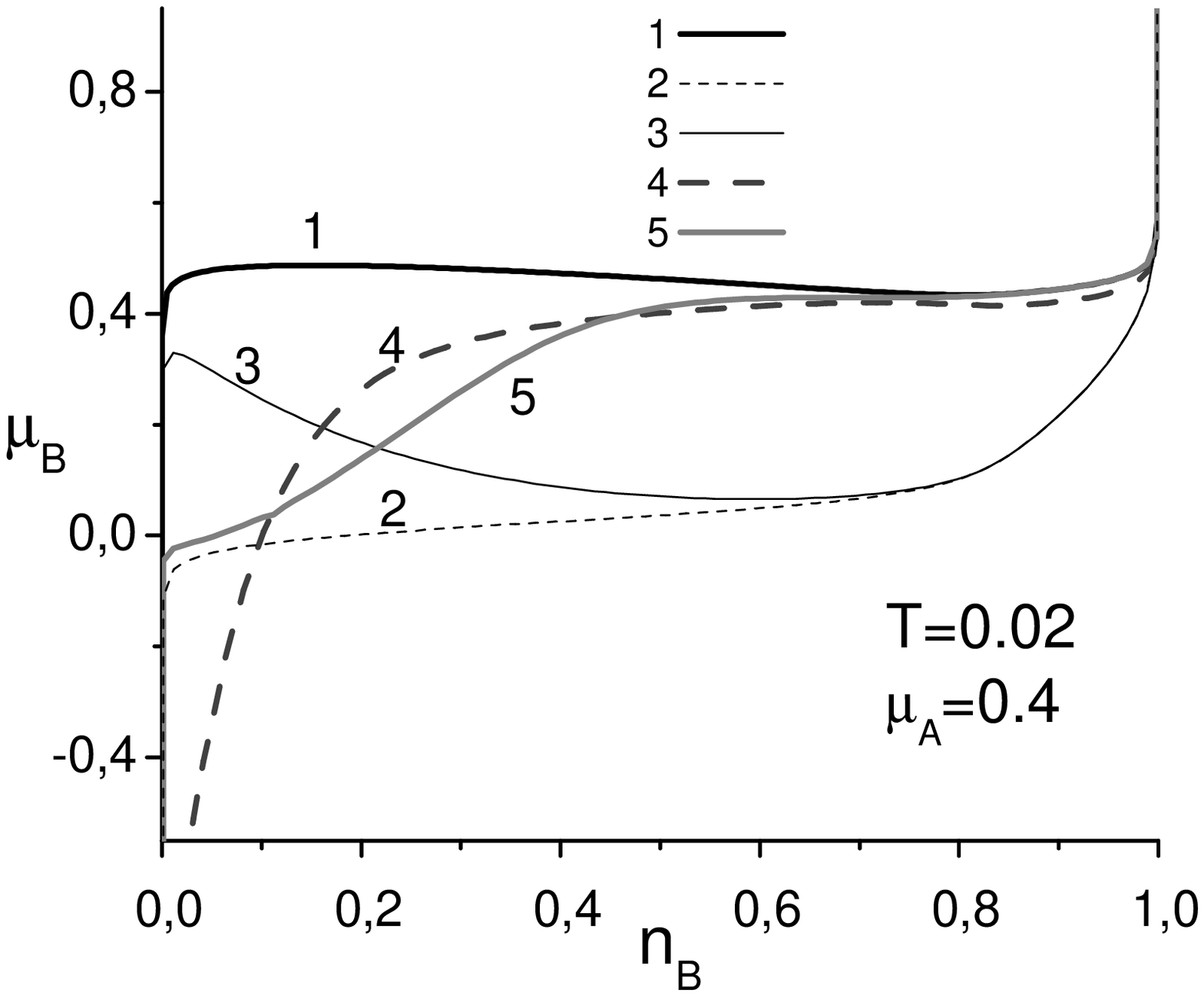}
\\
\includegraphics[width=0.32\textwidth]{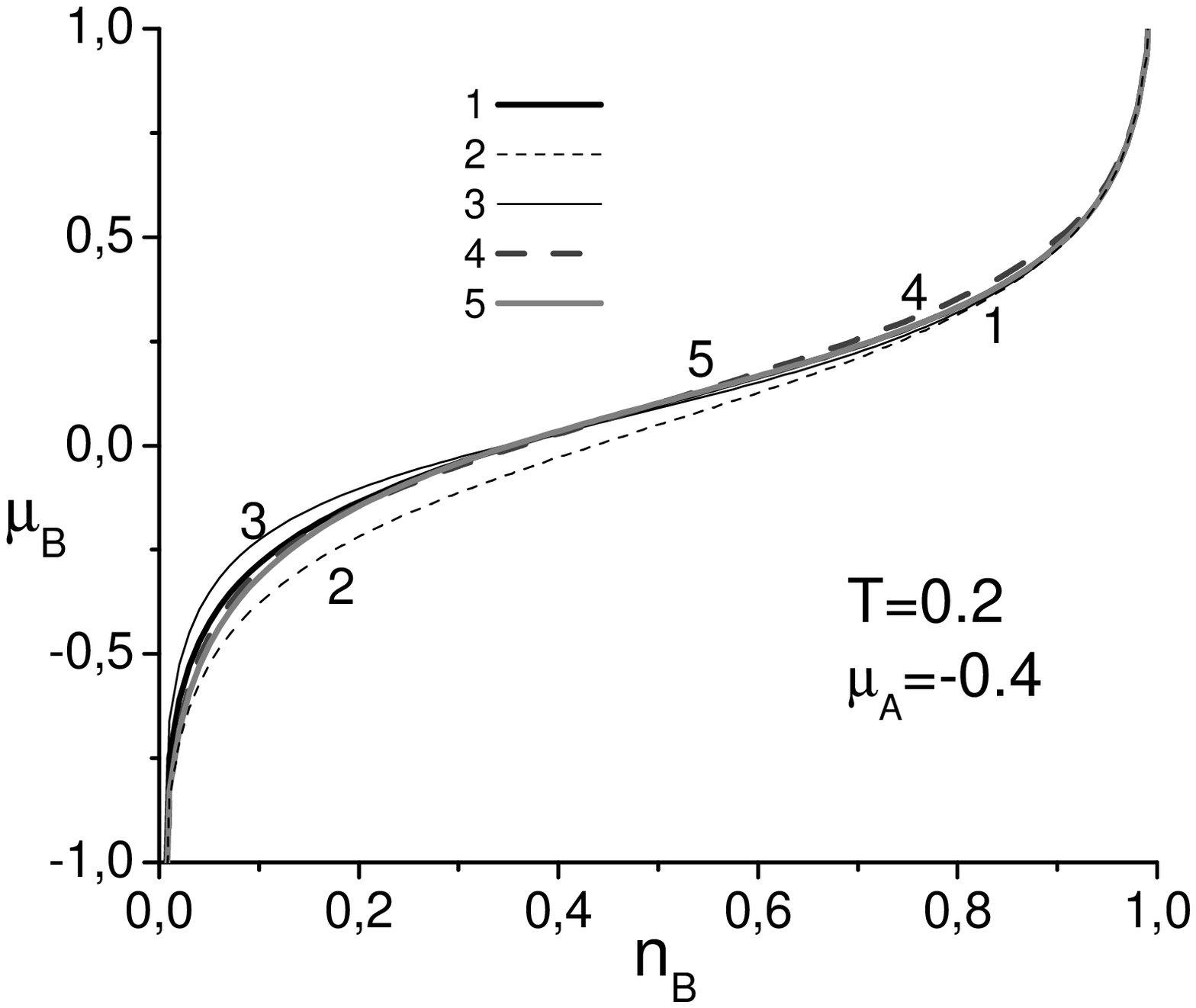}
\hfill
\includegraphics[width=0.32\textwidth]{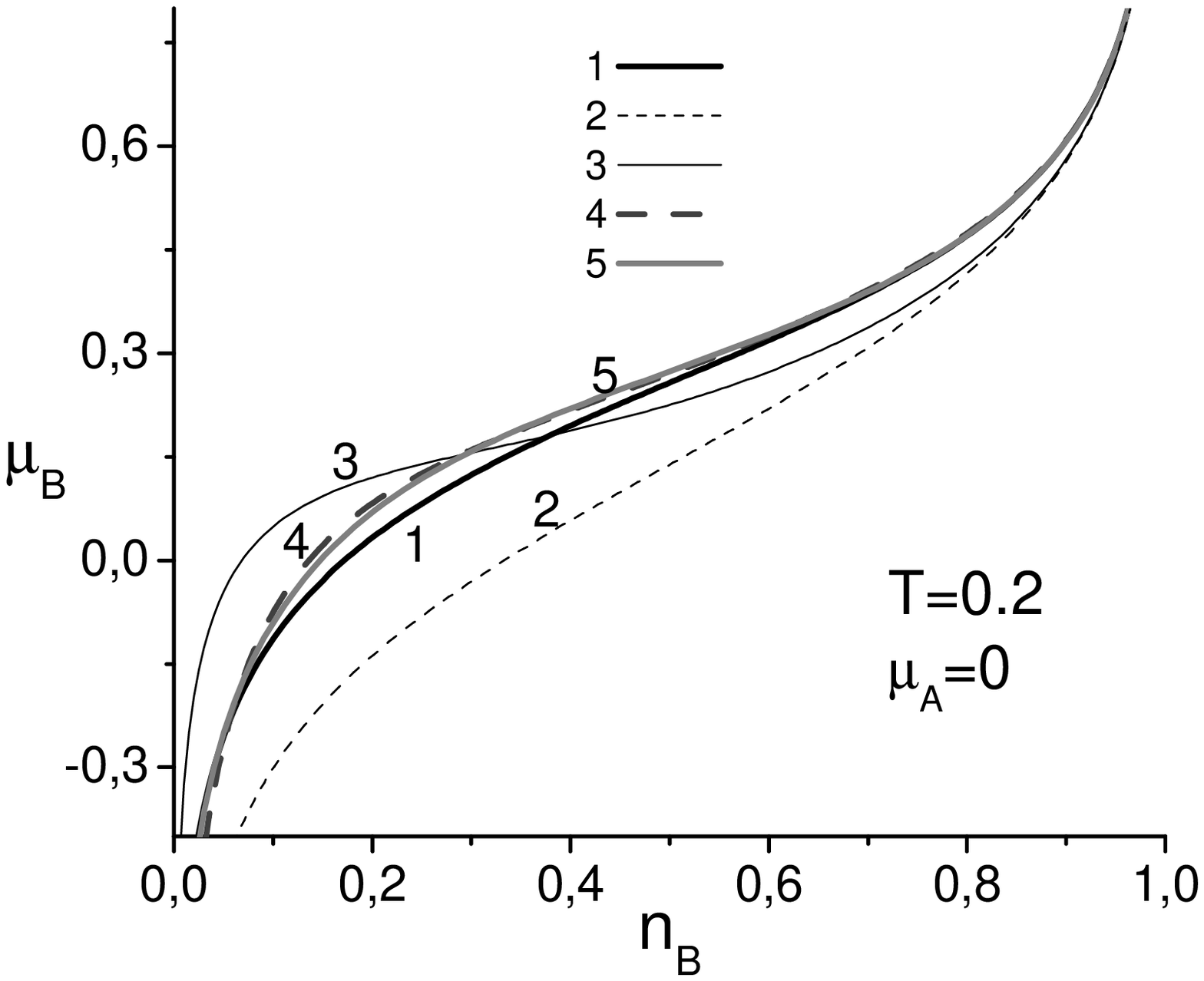}
\hfill
\includegraphics[width=0.32\textwidth]{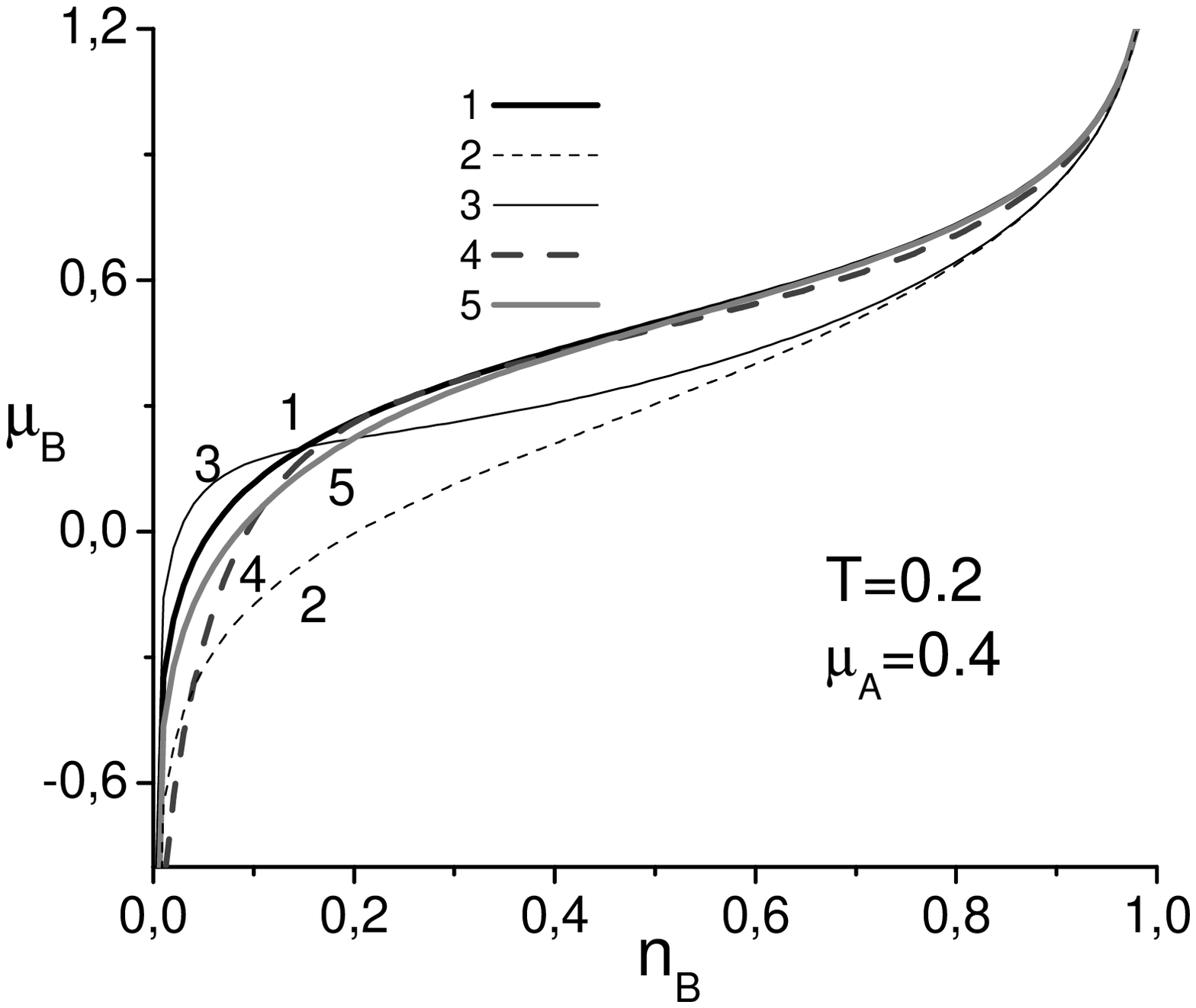}
\caption{ The $n_B$ dependence of $\mu_B$ in different approximations is
compared with the exact result obtained thermodynamically. The parameter
values: $W_A=1$, $W_B=0$. 1 -- exact result; 2 -- AA; 3 -- MAA; 4 -- H3;
5 -- GH3.}
\label{Fig_comp}
\end{figure}

\subsection{Approximate analytical approach}

Let us consider the results obtained at the Falicov-Kimball limit
of the considered model with $U \rightarrow +\infty$ in the case
when the above described approximate approach is  used instead of
the exact one (\ref{FFF1}). In the Falicov-Kimball limit, $J_B=0$
and it leads to $\varphi_A=0$, $R_A=0$. According to equation
(\ref{ss10}) the argument in the expression for $R_B (\omega)$
(\ref{ss4}) should be replaced by $\omega+\ri \varepsilon$.
Substituting the Green's function given by (\ref{GFK}) into this
expression we obtain
\begin{eqnarray}\label{AAAA1}
\lefteqn{ R_B(\omega+\ri \varepsilon) = -\frac{n_A+n_B}{2} J_A
(\omega+\ri \varepsilon+\mu_B-\mu_A)} \nonumber\\
&&{}+ \ri \frac{n_B-n_A}{2} \Imm  J_A (\omega+\ri
\varepsilon+\mu_B-\mu_A) \tanh\frac{\beta(\omega-\mu_A+\mu_B)}{2}
\nonumber\\
 &&{}+\frac{n_B-n_A}{4\pi} W_A^2(1-n_B) \,\mathrm{Vp} \int_{-1}^1
 \frac{\rd x \sqrt{1-x^2} \tanh \frac{\beta}{2}(xW_A\sqrt{1-n_B}-\mu_A)
 }{\omega+\mu_B-xW_A\sqrt{1-n_B}} {,}
\end{eqnarray}
where the principle value of the integral is taken. The similar
procedure for equation (\ref{ss6}) leads to
\begin{equation}
V\varphi_B=-\frac{W_A(1-n_B)^{3/2}}{2\pi} \,\mathrm{Vp}
\int_{-1}^1 \frac{\rd
x \sqrt{1-x^2}}{x}
          \Big[1+\re ^{\beta(x W_A \sqrt{1-n_B}-\mu_A)}\Big]^{-1}.
  \label{AAAA2}
\end{equation}
The $R_B(\omega)$ function provides in this approximation a
frequency dispersion of the Green's function of localized
particles and leads to the broadening of the corresponding density
of states into a band, (which is shifted with respect to the
initial level on the $-\frac{V}{1-n_A}\varphi_B$ value).

The dependence of the chemical potential on the concentration of
localized particles is calculated using expressions (\ref{ss10}),
(\ref{FFF12}), (\ref{AAAA1}), (\ref{AAAA2}). In
figure~\ref{Fig_comp} the approximations such as AA, MAA, GH3 and
H3 (in the last case the integral terms that describe the
processes of scattering by boson excitations are neglected in the
expressions for $R_\sigma$) are compared with the exact results
obtained thermodynamically. The simple AA approximation cannot
describe phase transitions because it gives a monotonically
increasing dependence $\mu_B=\mu_B(n_B)$. MAA and AA
approximations can describe the thermodynamics of the system
at high temperatures or at a negative value of $\mu_A$ (small
concentration of moving particles). The H3 approximation gives
better results for high temperatures but not for the low ones. The
best results are given by the GH3 approximation for high
temperatures, as well as for low temperatures in the cases of a
nearly full or empty band of the moving particles (and high values
of $n_B$).

\begin{figure}
\includegraphics[width=0.32\textwidth]{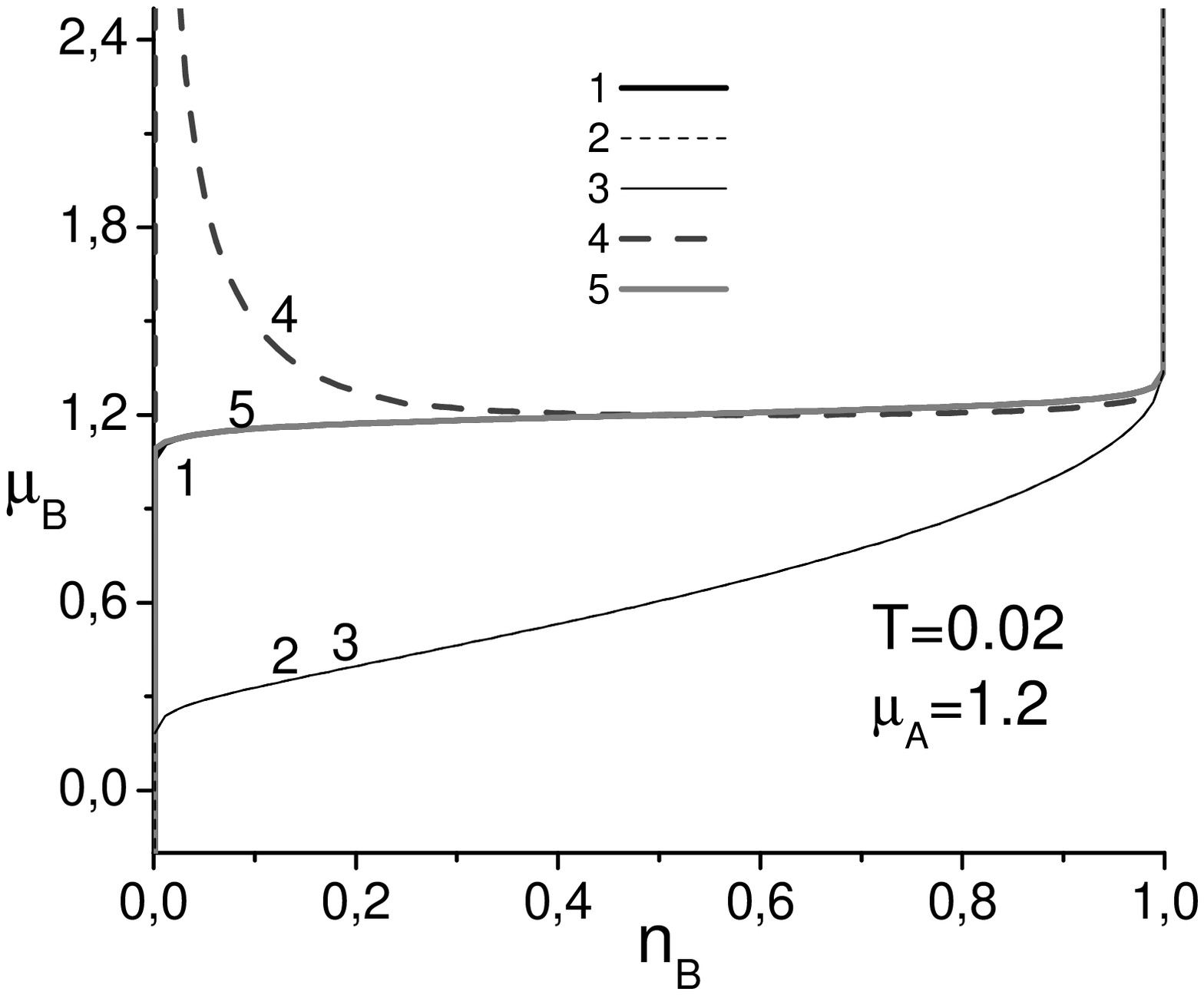}
\hfill
\includegraphics[width=0.32\textwidth]{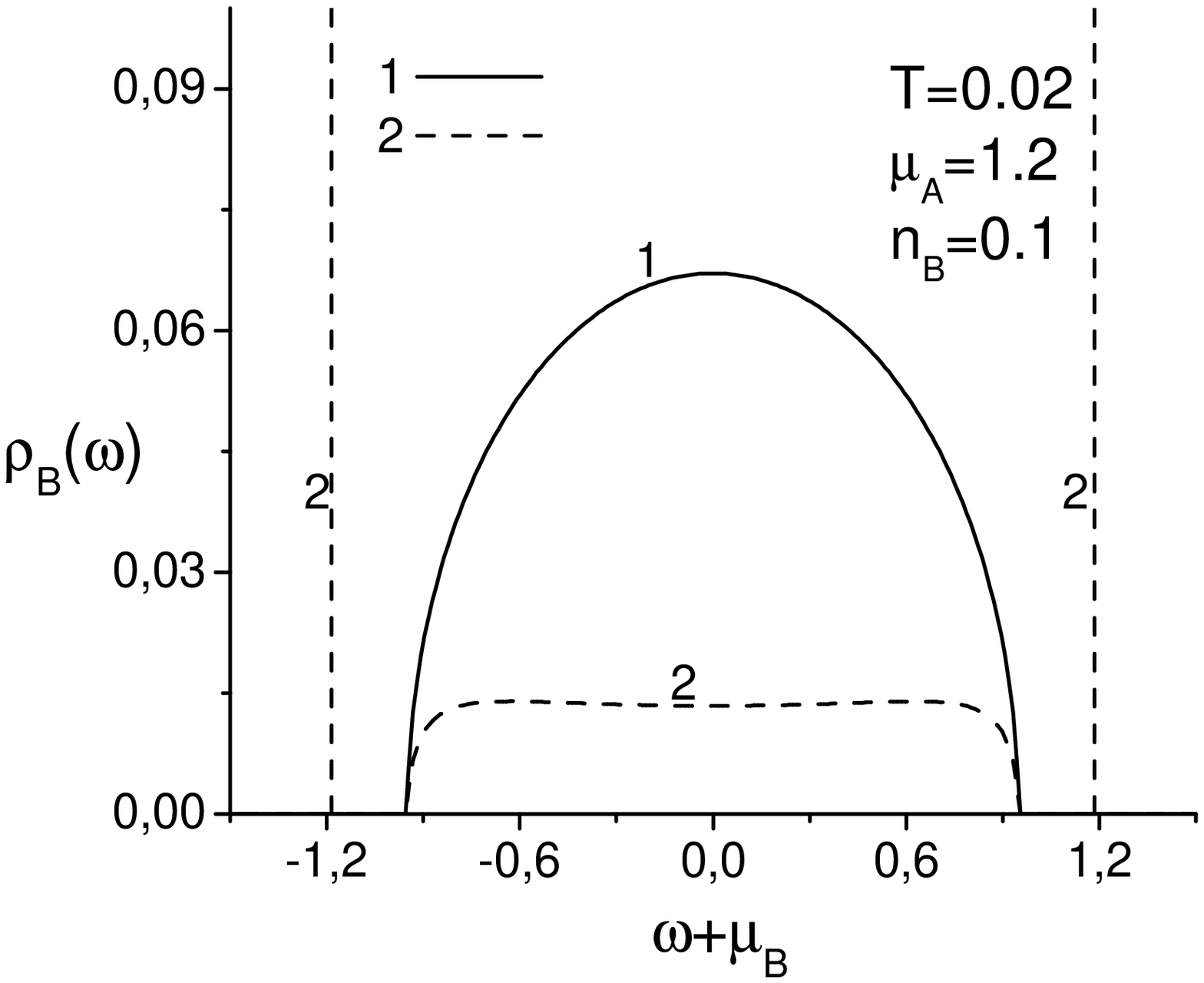}
\hfill
\includegraphics[width=0.32\textwidth]{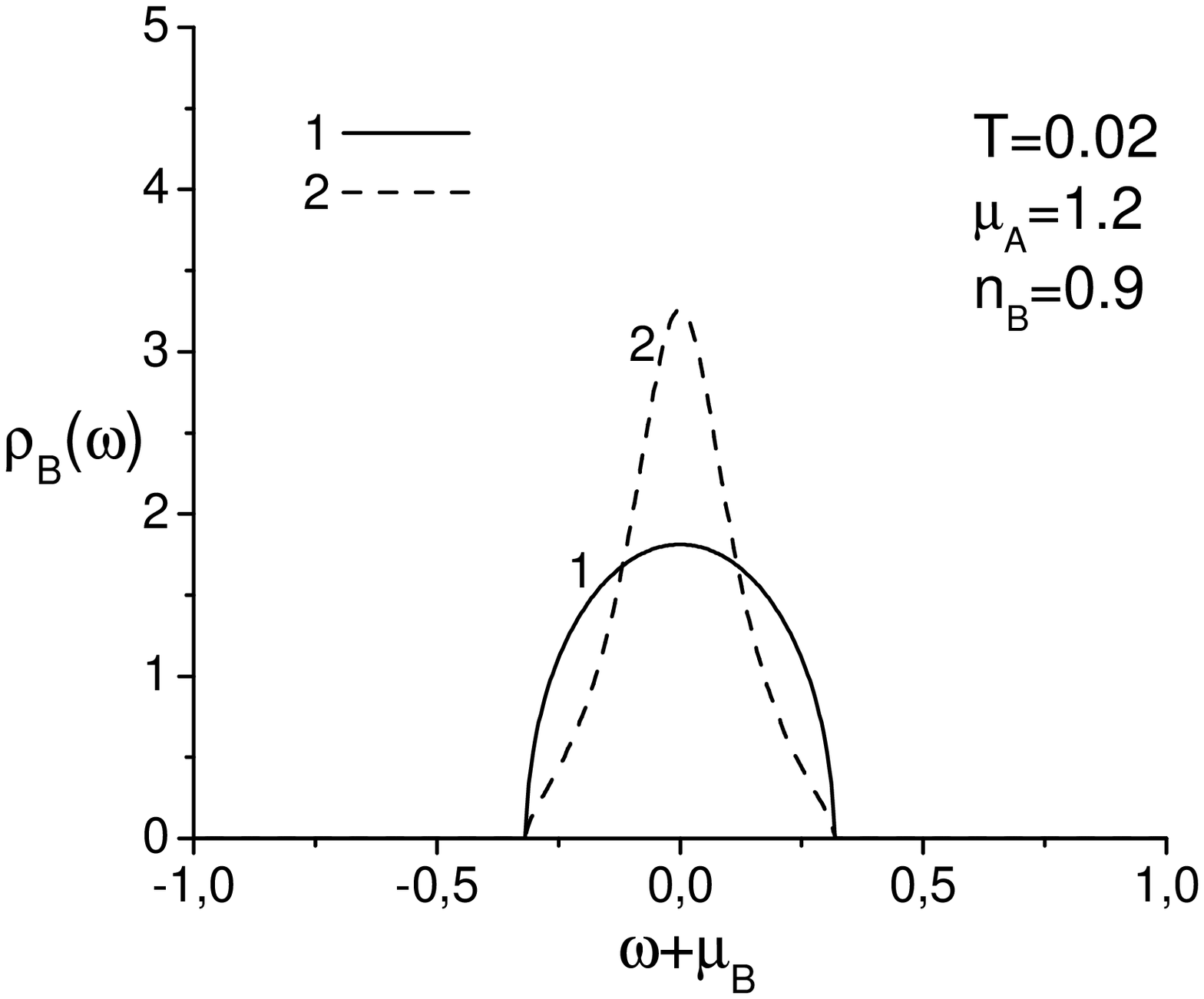}
\\
\includegraphics[width=0.32\textwidth]{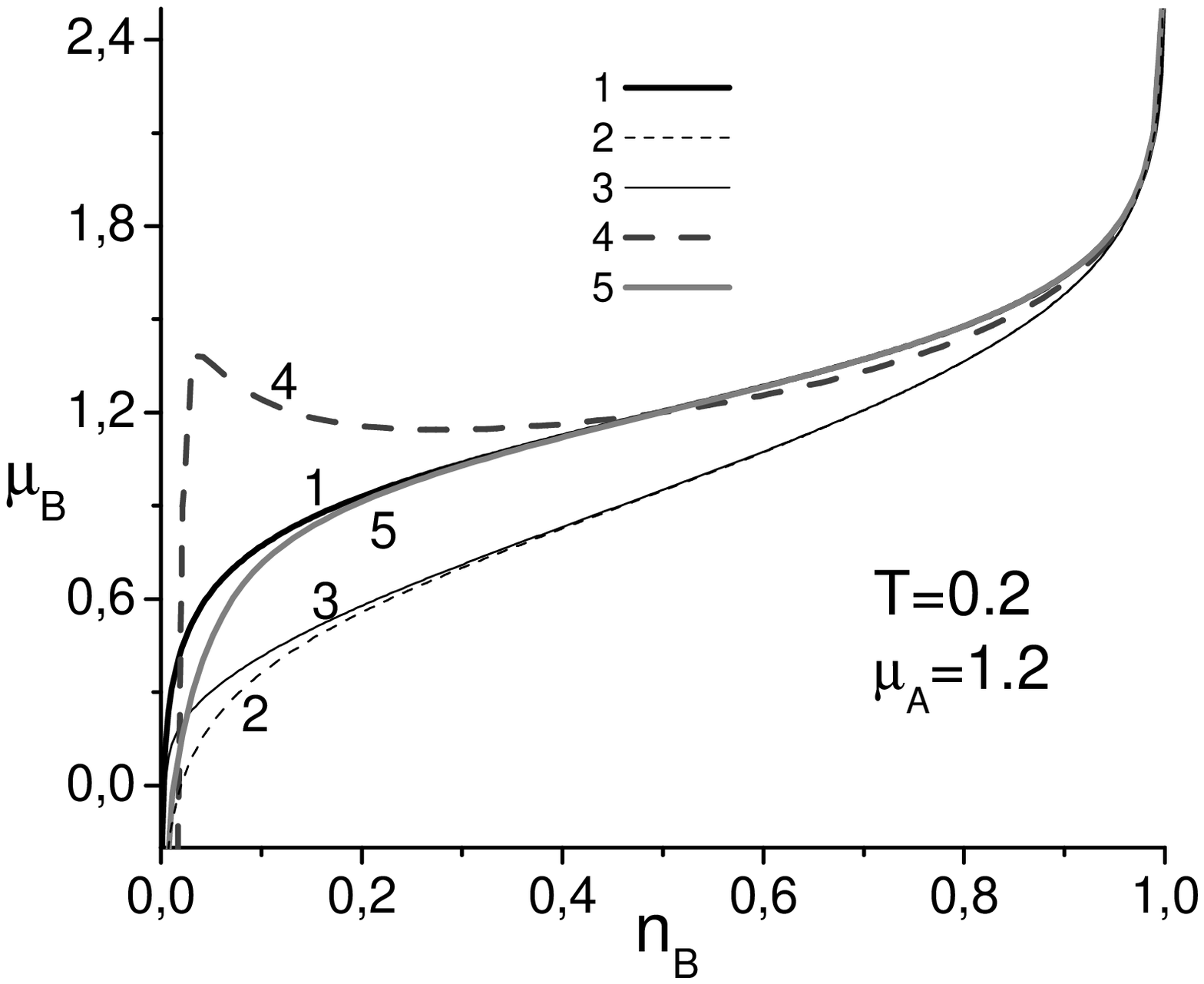}
\hfill
\includegraphics[width=0.32\textwidth]{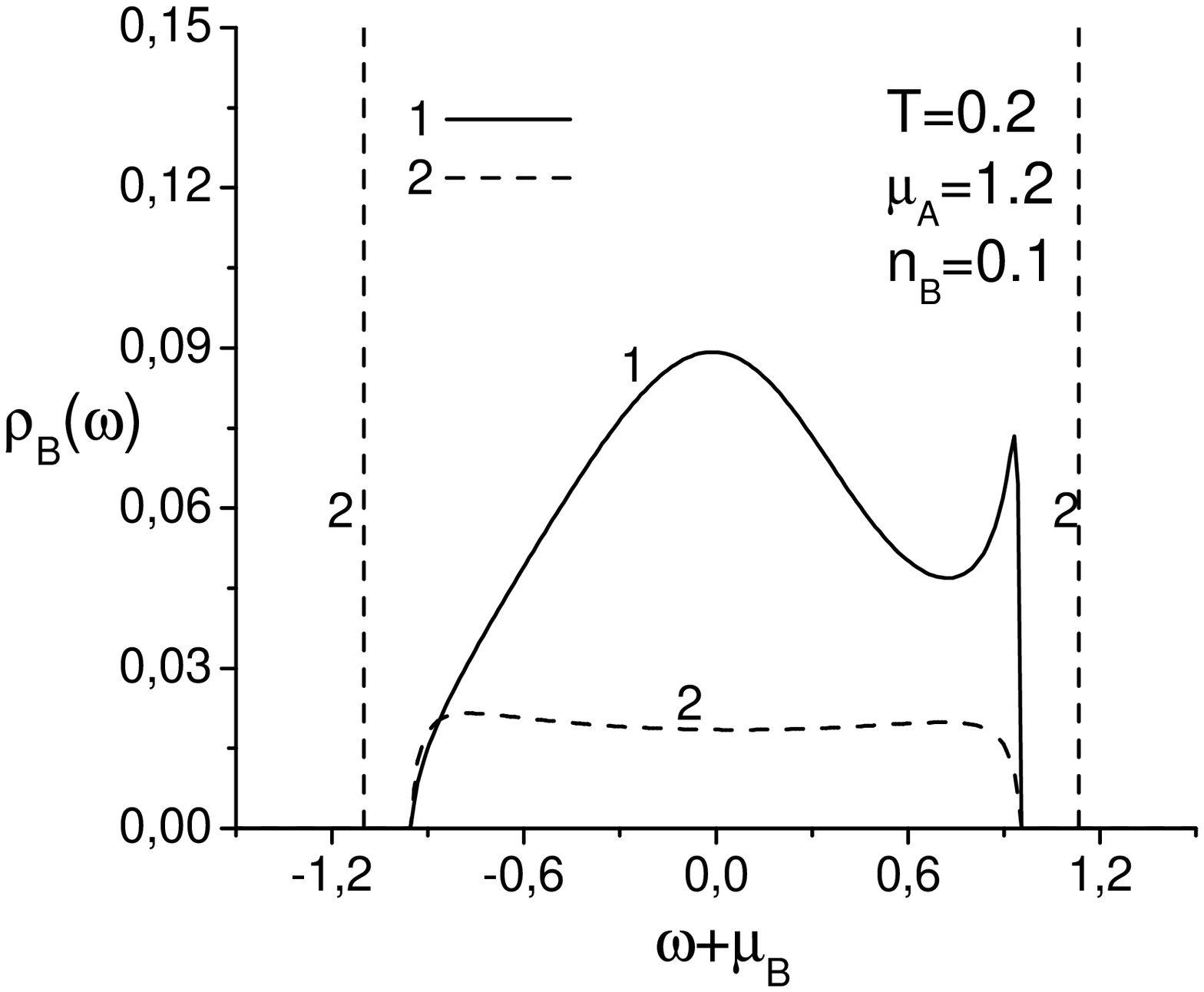}
\hfill
\includegraphics[width=0.32\textwidth]{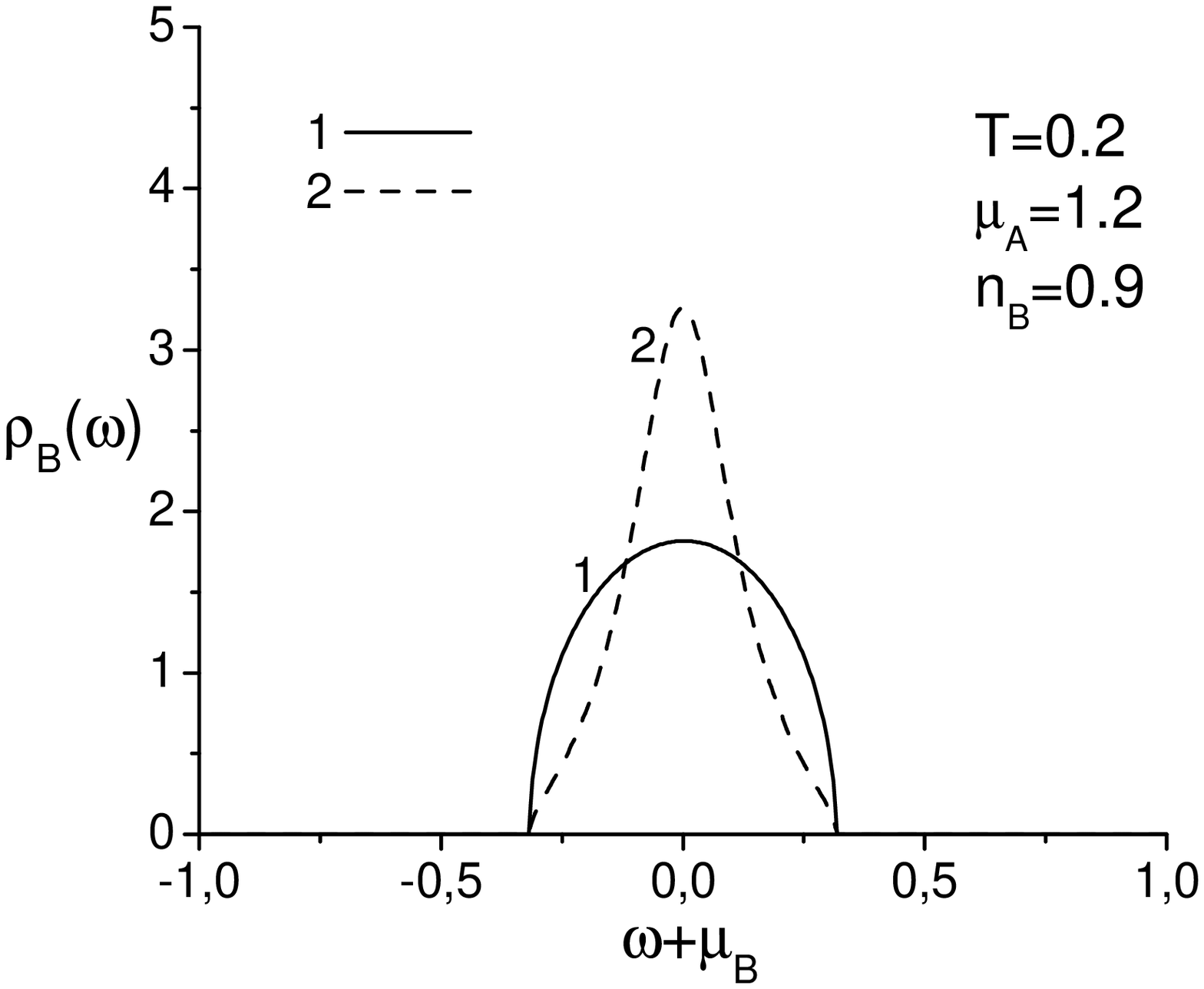}

\caption{$\mu_B$ as a function of $n_B$ in different
approximations (1 -- exact result; 2 -- AA; 3 -- MAA; 4 -- H3; 5
-- GH3) and corresponding densities of states $\rho_B$ of
localized particles (1 -- GH3; 2 -- H3).}\label{Fig_comp2}
\end{figure}

In figure~\ref{Fig_comp2} the case of a nearly full band of the
moving particles is shown. The $\mu_B=\mu_B(n_B)$ dependence is
illustrated together with the corresponding curves for density of
states of localized particles in H3 and GH3 approximations:
\begin{equation}
 \rho_B(\omega)=-\frac{1}{\pi} \Imm  G_B^{(a)}(\omega+\ri \varepsilon).
\end{equation}
One can see that this band can have a complicated structure or
even be split in the H3 approximation, but it does not correspond
to the satisfactory $\mu_B(U_B)$ dependence. In general, there is
a considerable difference between densities of states
$\rho_B(\omega)$ given by the H3 and GH3 approximations, but the
band edges are determined properly even in the H3 approximation.

\begin{figure}
\includegraphics[width=0.32\textwidth]{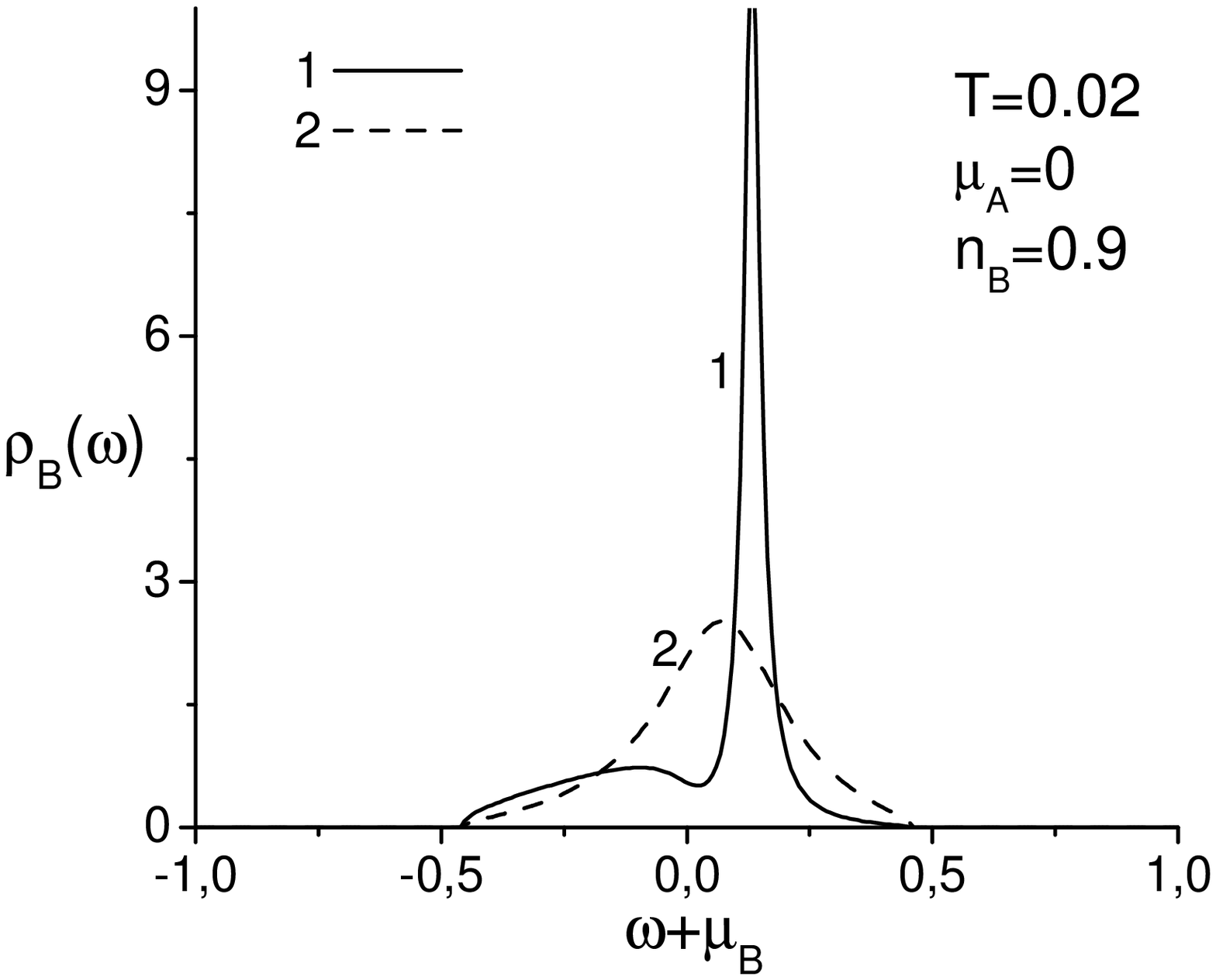}
\hfill
\includegraphics[width=0.32\textwidth]{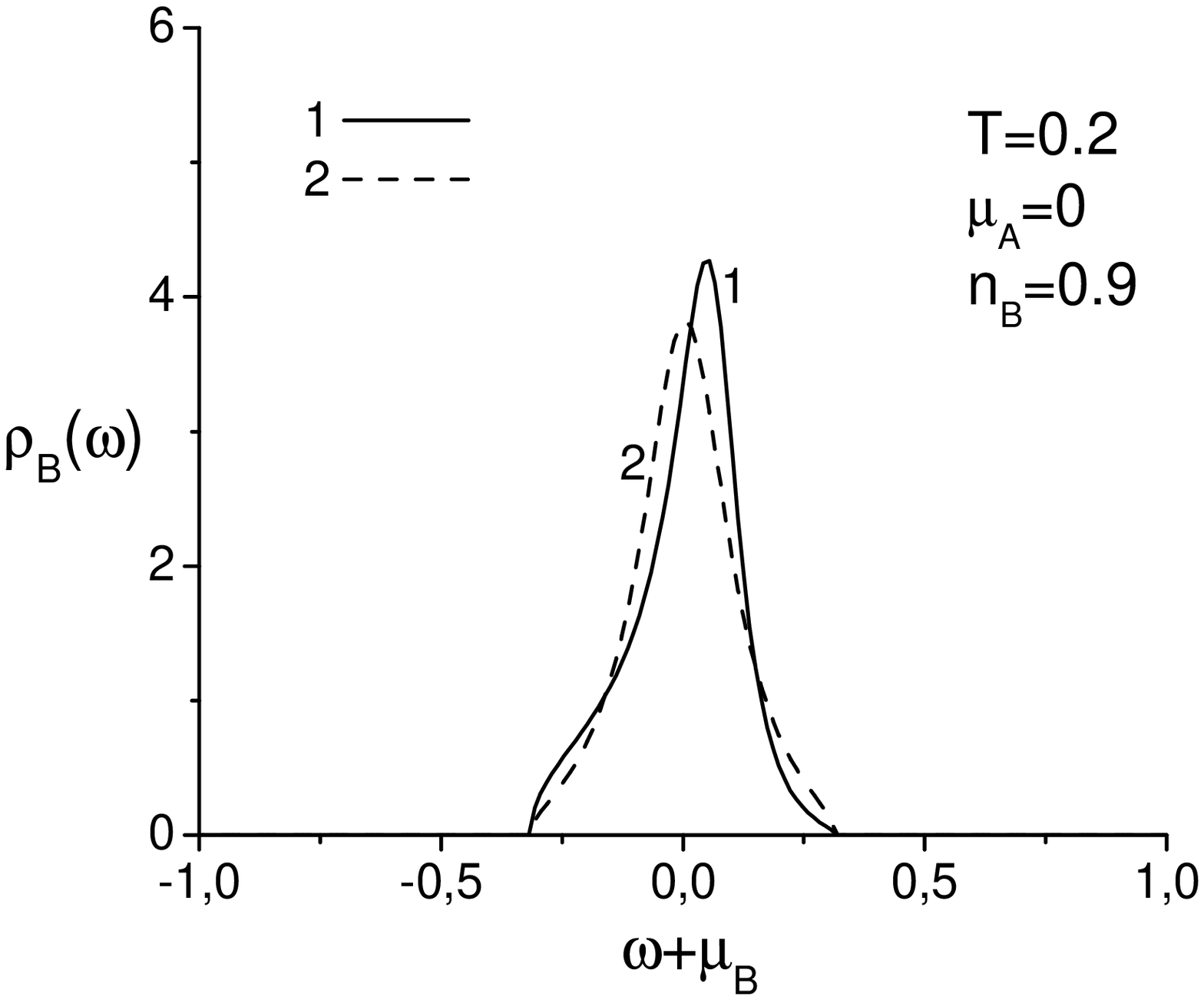}
\hfill
\includegraphics[width=0.32\textwidth]{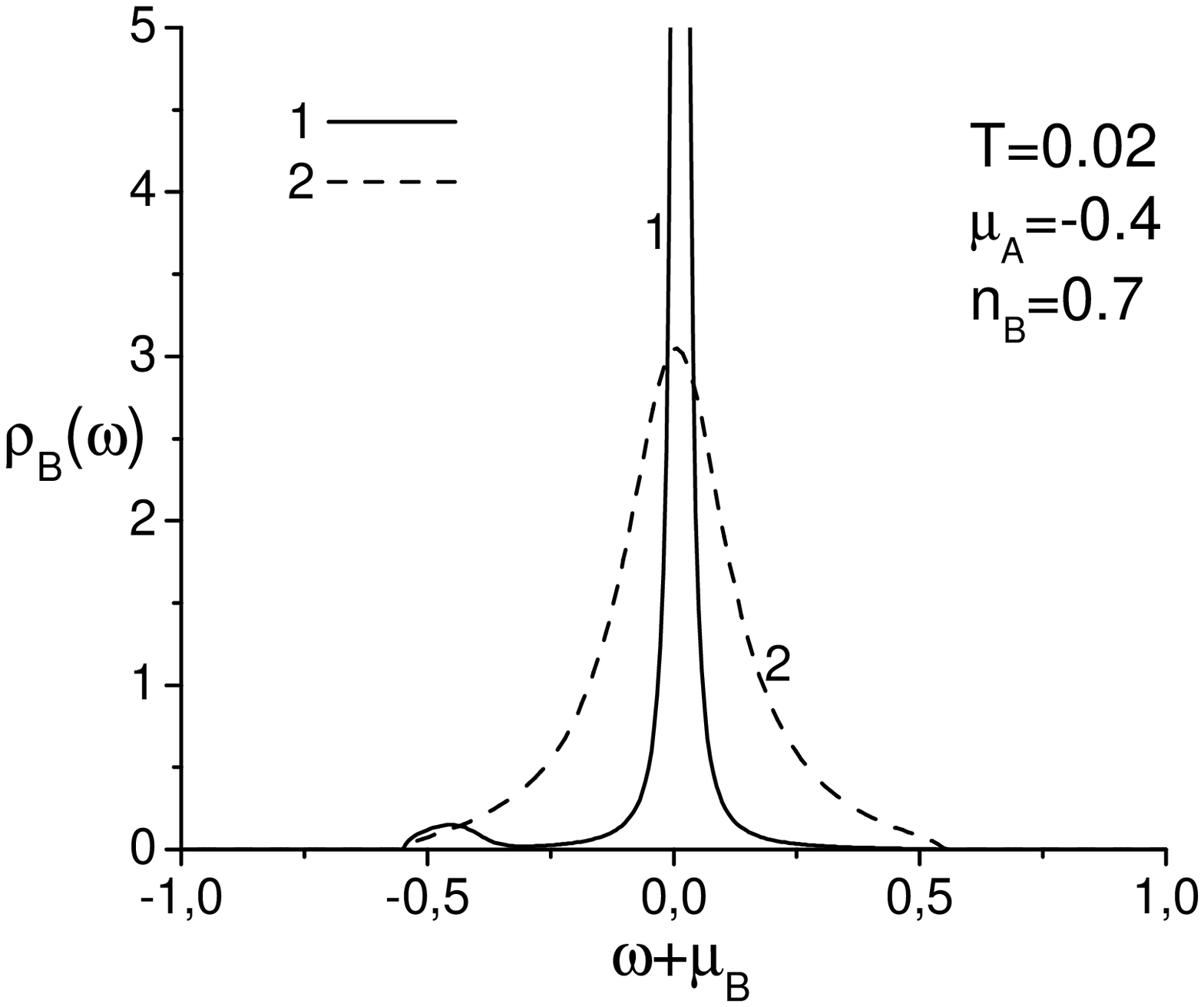}
\\
\includegraphics[width=0.32\textwidth]{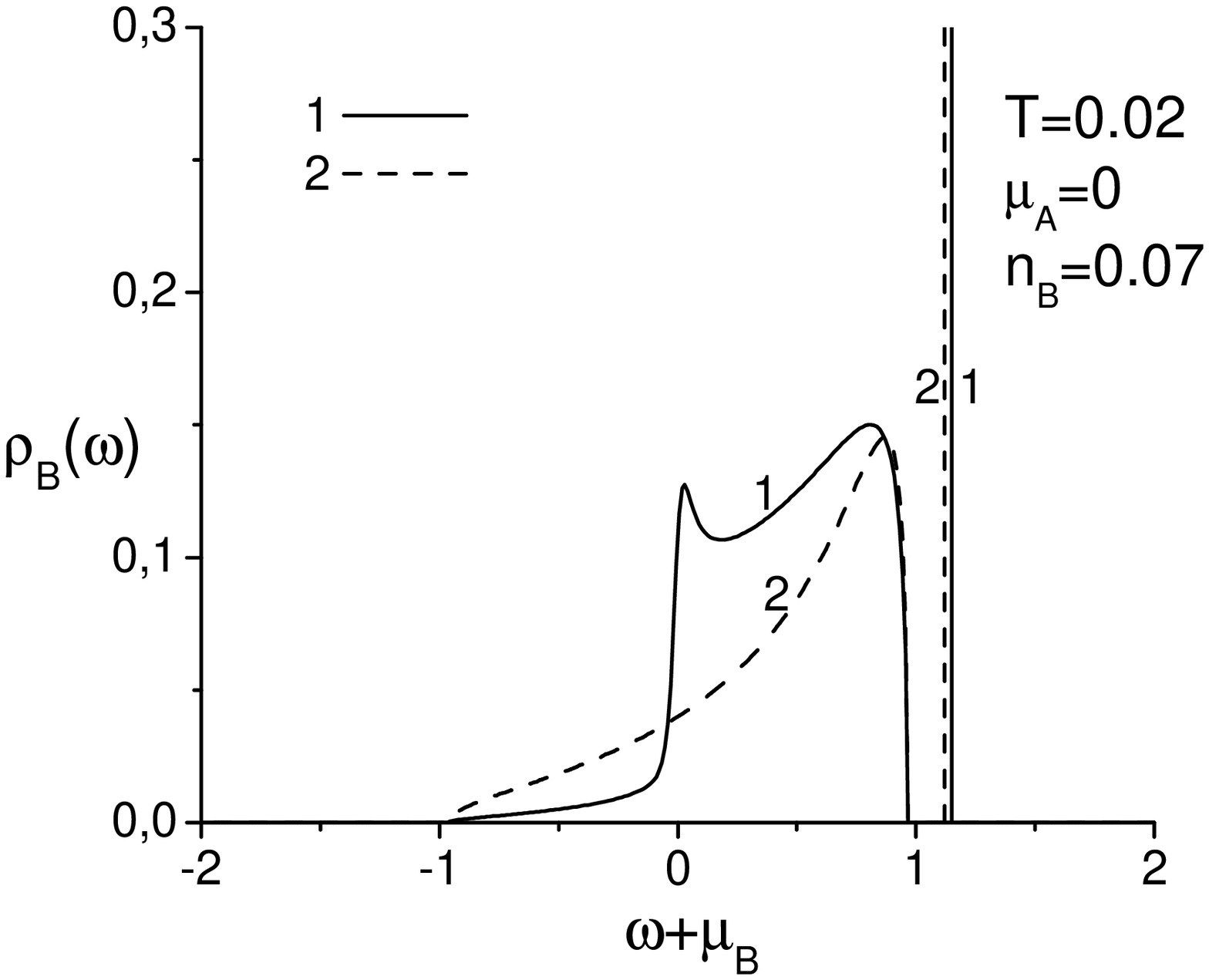}
\hfill
\includegraphics[width=0.32\textwidth]{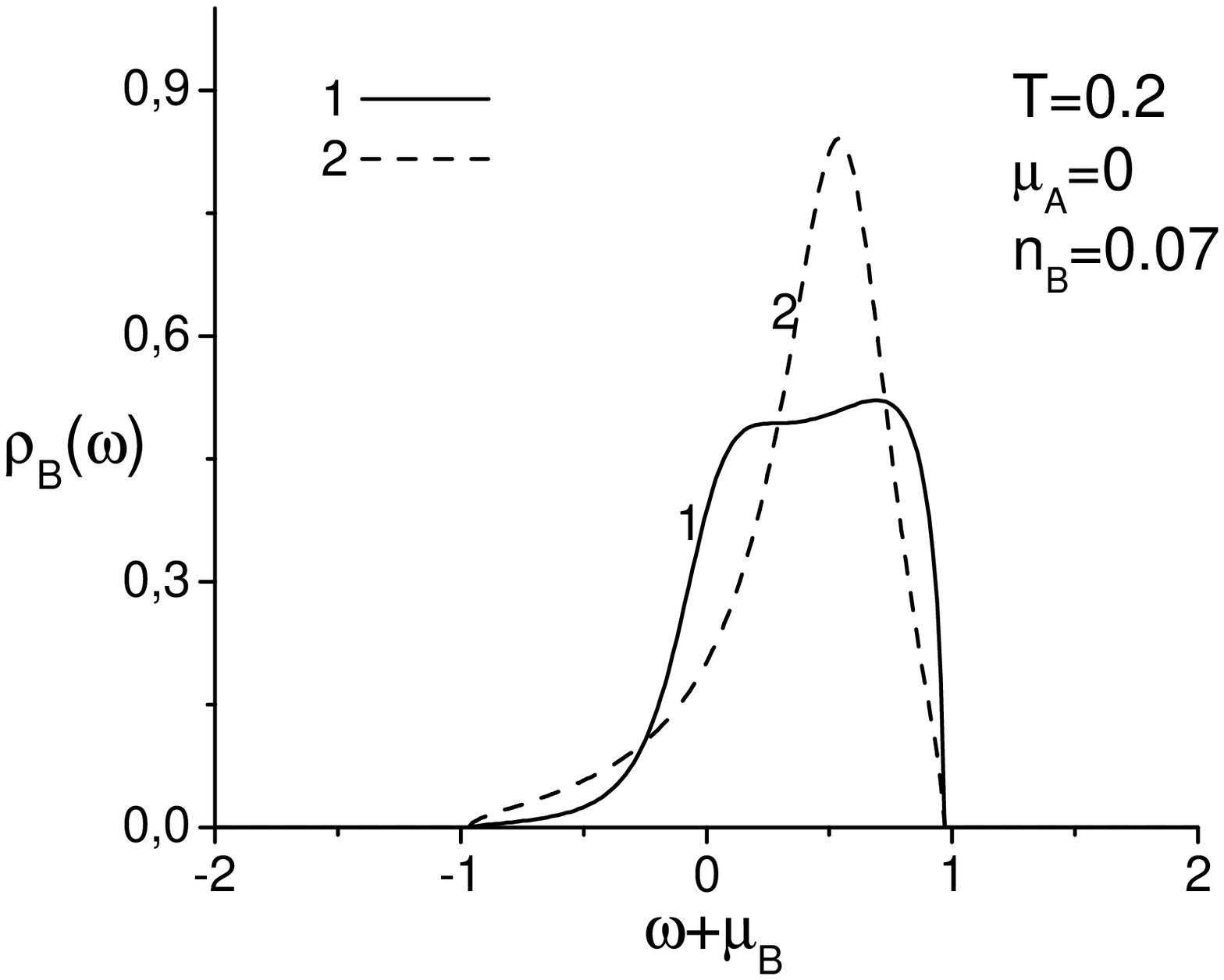}
\hfill
\includegraphics[width=0.32\textwidth]{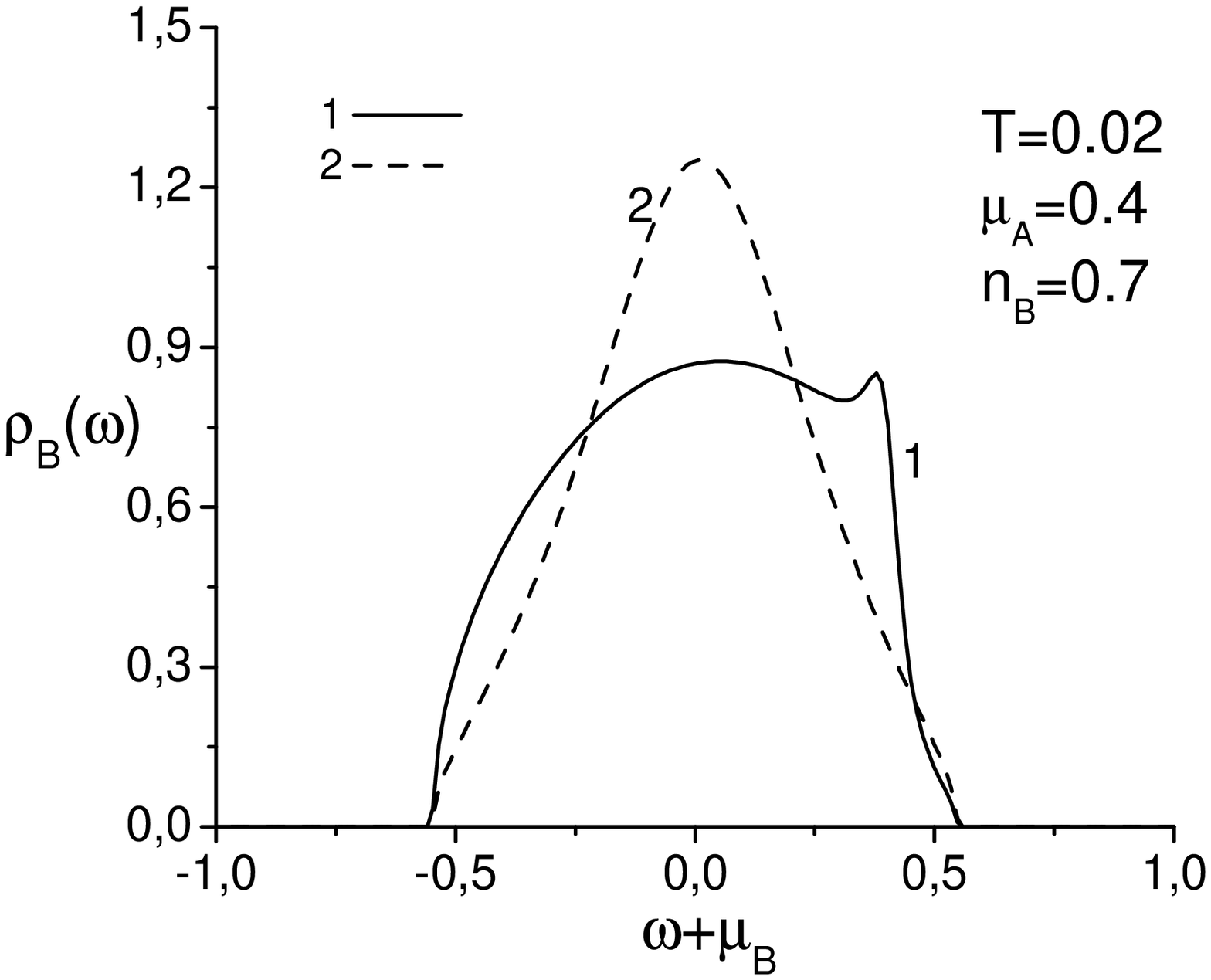}
\caption{The approximate densities of states of localized
particles (1~--~GH3; 2~--~H3). }\label{Fig_comp3}
\end{figure}

In figure~\ref{Fig_comp3} the densities of states of localized
particles are presented for different temperatures and particle
concentrations. In the cases corresponding to the above-mentioned
criteria ((i)~high $T$; (ii)~low $T$ and chemical potential
$\mu_A$ close to the edges of the band of the moving particles),
the obtained densities of states $\rho_B$ can be considered as
close to the real ones due to a satisfactory description of
the $\mu_B(n_B)$ dependences.

Calculation of densities of states of localized particles in the
Falicov-Kimball model had been done earlier within DMFT in
\cite{brur,Freericks4}. The algorithm which leads to exact results
is rather complicated. The plots obtained in
\cite{brur,Freericks4} correspond to finite values of $U$ and to
the symmetrical case of half-filling (the hypercubic lattice with
Gaussian density of states was considered). The band of localized
particles is split and consists of two subbands for $U$ larger
than some critical value ($U \gtrsim 5~W_A$); for $U\rightarrow
+\infty$, the only lower subband remains. Our results correspond
by the general shape of the curves to the above-mentioned results
in the cases illustrated in figure~\ref{Fig_comp2} for $T=0.02\;\
0.2$; $\mu_A=1.2$; $n_B=0.9$ and in figure~\ref{Fig_comp3} for
$T=0.2$; $\mu_A=0$; $n_B=0.9$ within the GH3 approximation (in
these cases, there is the best agreement with the exact results
for $\mu_B(n_B)$ dependences). Quantitative comparison of
densities of states of localized particles will be possible after
doing calculations with nonperturbed density of state like in
\cite{brur,Freericks4} and for finite values of $U$.

\section{Conclusions}

Application of the approximate analytical method of solving the
effective single-site problem within DMFT gives a basic set of
equations for evaluating single-particle Green's functions and for
investigating an energy spectrum of the asymmetric Hubbard model
describing the system of particles of two types with different
transfer parameters ($t^A\neq t^B$) on a lattice.

The limiting case $U=\infty$ (exclusion of a double occupation
of a site) and $t^B\rightarrow 0$ (the case of the infinitesimally
small mobility of particles of one of the type, when the model becomes
the Falicov-Kimball model) is considered.

Phase transitions are investigated in thermodynamic regimes
of the fixed values of the following quantities:
(i)~$\mu_A$, $\mu_B$; (ii)~$\mu_A$, $n_B$; or (iii)~$\mu_B$, $n_A$.
Phase diagrams describing coexistence curves of homogeneous phases
with different $n_A$ and $n_B$ values and phase separation regions
are obtained. These results extending a space of thermodynamic
parameters complement the existing information about the
thermodynamics of the Falicov-Kimball model in the $U\rightarrow
+\infty$ limit.

The set of self-consistent equations, which relates the
concentrations (chemical potentials) of particles of $A$ and $B$
types, coherent potentials $J_{A,B}(\omega)$ and shift constants
of energy spectrum $\varphi_{A,B}$, is solved in two cases:
(i)~with an application of the exact thermodynamic relations for
the Falicov-Kimball model to derive an equation for concentration
$n_B$ of localized particles; (ii)~by using the approximate
analytical scheme in DMFT. Calculation of $\mu_B(n_B)$ dependences
being done by the two mentioned methods allows us to investigate
an applicability of this approximate approach. The results closest
to the exact ones are obtained within the GH3 approximation for
high temperatures, as well as for low temperatures in the cases of
nearly full or nearly empty band of the moving particles and for
high values of $n_B$.

Densities of states $\rho_B(\omega)$ of localized particles are
calculated using the approximate DMFT scheme. The results obtained
in the GH3 approximation can be considered as the closest to the
real ones, when a satisfactory dependence of $\mu_B(n_B)$ is
achieved. Extending this approach to the cases of low temperatures
and intermediate band filling requires an improvement of the
applied scheme, namely, a more precise (probably self-consistent)
way of taking into account the processes of scattering by boson
excitations for calculating the $R_B(\omega)$ function which
determines the energy spectrum of localized particles.

\section*{Acknowledgements}

This work was partially supported by the Fundamental Researches
Fund of the Ministry of Ukraine of Science and Education
(Project No.~02.07/266).

\label{last@page}

\end{document}